\documentclass[sn-mathphys-num]{sn-jnl}

\usepackage{graphicx}%
\usepackage{multirow}%
\usepackage{amsmath,amssymb,amsfonts}%
\usepackage{amsthm}%
\usepackage{mathrsfs}%
\usepackage[title]{appendix}%
\usepackage{xcolor}%
\usepackage{textcomp}%
\usepackage{manyfoot}%
\usepackage{booktabs}%
\usepackage{algorithm}%
\usepackage{algorithmicx}%
\usepackage{algpseudocode}%
\usepackage{listings}%

\usepackage{float} 
\usepackage{graphicx}
\usepackage{arydshln}
\usepackage{mathtools}

\DeclarePairedDelimiter\floor{\lfloor}{\rfloor}
\usepackage{multicol}
\usepackage[nice]{nicefrac}
\usepackage{tabularray}
\usepackage{siunitx}
\usepackage[left]{lineno}
\usepackage{changepage}  

\usepackage{geometry}
\newgeometry{bottom=2.5cm, right=2.5cm, left=2.5cm, top=2.5cm}
\usepackage{fancyhdr}

\begin{document}


\fancypagestyle{firstpage}{
  \fancyhf{} 
  \fancyhead[L]{\textbf{Received: 20 Feb 2024, Accepted: 13 June 2024, Published: 01 July 2024.\\
  Nonlinear Dynamics\\
https://doi.org/10.1007/s11071-024-09890-4}}
  \renewcommand{\headrulewidth}{0pt}
}
\thispagestyle{firstpage} 

\title[title]{
System identification based on characteristic curves: a mathematical connection between power series and Fourier analysis for first-order nonlinear systems}

\author*[1,2]{\fnm{Federico J.} \sur{Gonzalez}}\email{fgonzalez@ifir-conicet.gov.ar}

\affil[1]{\orgname{Instituto de F\'isica Rosario (IFIR-CONICET)}, \orgaddress{\street{Bv. 27 de Febrero 210 Bis}, \city{Rosario}, \postcode{S2000EZP}, \country{Argentina}}}

\affil[2]{\orgname{Facultad de Ciencias Exactas, Ingenier\'ia y Agrimensura (UNR)}, \orgaddress{\street{Av. Pellegrini 250}, \city{Rosario}, \postcode{S2000BTP}, \country{Argentina}}}


\abstract{
Recently, the \textbf{s}inosoidal \textbf{o}utput \textbf{r}esponse in \textbf{p}ower \textbf{s}eries (SORPS) formalism was presented for system identification and simulation. Based on the concept of \textbf{c}haracteristic \textbf{c}urves (CCs), it establishes a mathematical connection between power series and Fourier series for a first-order nonlinear system [F. J. Gonzalez, Sci. Rep. 13, 1955, (2023)]. 
However, the system identification procedure discussed there, based on \textbf{f}ast \textbf{F}ourier \textbf{t}ransform (FFT), presents the limitations of requiring a sinusoidal single tone for the dynamical variable and equally spaced time steps for the input-output \textbf{d}ata\textbf{s}et (DS). 
These limitations are here addressed by introducing a different approach: we use a power series-based model (referred to as model 1) for system modeling instead of FFT, where two hyperparameters $\hat{A}_0$ and $\hat{A}_1$ are optimally defined depending on the DS. 
Subsequently, two additional models are obtained from parameters obtained in model 1: another power series-based model (model 2) and a Fourier analysis-based model (model 3). These models are useful for comparing parameters obtained from different DSs. 
Through an illustrative example, we show that while the predicted values from the models are the same due to a mathematical equivalence, the parameters obtained for each model vary to a greater or lesser extent depending on the DS used for system estimation.  
Hence, the parameters of the Fourier analysis-based model exhibit notably less variation compared to those of the power series-based model, highlighting the reliability of using the Fourier analysis-based model for comparing model parameters obtained from different DSs. 
Overall, this work expands the applicability of the SORPS formalism to system identification from arbitrary input-output data and represents a groundbreaking contribution relying on the concept of CCs, which can be straightforwardly applied to higher-order nonlinear systems.  
The method of CCs can be considered as complementary to the commonly used approach (such as NARMAX-models and sparse regression techniques) that emphasizes the estimation of the individual parameter values of the model. Instead, the CCs-based methods emphasize the computation of the CCs as a whole. CCs-based models present the advantages that the system identification is uniquely defined, and that it can be applied for any system without additional algebraic operations.
Thus, the parsimonious principle defined by the NARMAX-philosophy is extended from the concept of a model with as few parameters as possible to the concept of finding the lowest model order that correctly describe the input-output data.  
This opens up a wide variety of potential applications, covering areas such as vibration analysis, structural dynamics, viscoelastic materials, design and modeling of nonlinear electric circuits, voltammetry techniques in electrochemistry, structural health monitoring, and fault diagnosis.
}


\keywords{nonlinear system identification, Fourier analysis, characteristic curves, parsimonious model, frequency domain, least-squares regression method}



\maketitle

\section{Introduction}\label{sec:introduction}
The field of nonlinear system analysis and identification is extensive, with significant contributions dating back to the development of Volterra series\cite{Volterra1887,Volterra1958} in the 1890s and the subsequent advancements of Wiener series\cite{Wiener1958,Wiener1964} from the 1940s to the 1960s. 
While these methods remain actively studied as methods of analysis\cite{Marmarelis1978,doyle2002,PeytonJones2017,PeytonJones2018}, their application in the analysis of highly nonlinear systems and for system identification continues to be challenging, primarily due to issues such as convergence, stability, and overparameterization\cite{Favier2012,CHENG2017340,Annabestani2019,dePaula2019,Skyvulstad2023}.  
In addition, various models were developed for identifying systems that can be decomposed into linear and nonlinear subsystems, such as the Wiener, Hammerstein , and Wiener-Hammerstein models in the 2000s (see Refs.~\cite{Gomez2004,WILLS201370,tiels2014wiener,Cheng2016,Kazemi2017,Hammar2019} and references therein). While still applicable in real-world scenarios, their main drawback is their limited us, as they are designed for specific types of nonlinear systems.



Simultaneously, throughout the last century, Fourier analysis has emerged as a powerful technique for studying linear systems. It explores how a function can be represented as a superposition of trigonometric functions, providing a representation of the system in the reciprocal space. In the case of time domain variables, this reciprocal space corresponds to the frequency domain, while for spatial domain variables, it is the momentum space. 

For \textbf{l}inear \textbf{t}ime \textbf{i}nvariant (LTI) systems, the superposition principle holds, implying that the response of the system to a single sinusoidal tone is also a sinusoidal tone with the same frequency but eventually differing by an amplitude scaling and a phase shifting. 
Primarily, this linearity property has positioned Fourier analysis as the predominant theory for studying linear systems. 
Furthermore, the use of reciprocal space representation transforms linear differential equations into algebraic equations, simplifying the solution process.
As a consequence, Fourier analysis has experienced extensive development over the last century through various approaches and has been mathematically formalized under the name of harmonic analysis. Nowadays, Fourier analysis stands as an indispensable tool in practically all scientific and engineering areas for studying linear systems\cite{Brigham1988,Oppenheim1999,Proakis20074th,Stoica2005,Kammler2008,Pintelon2012}.  
However, the majority of systems are naturally nonlinear. 
In these systems, the superposition principle is not valid, and as a result, the response to a single sinusoidal tone becomes much more complex, giving rise to interesting phenomena such as superharmonics (integer multiples of the fundamental frequency), subharmonics (integer fractions of the fundamental frequency), interharmonics (non-integer multiples of the fundamental frequency), and chaos (characterized by aperiodic and unpredictable behavior).

Considering the widespread success of Fourier analysis in studying linear systems and its role as a natural starting point for exploring nonlinear systems, it becomes compelling to precisely determine how Fourier analysis and other related reciprocal space-based techniques can be applied to the study of nonlinear systems.
In this context, nearly all methods developed for studying nonlinear systems have been investigated for their representations in reciprocal space.


Historically, contributions from Poincar{\'e}\cite{Poincare1957} in the 1890s, Van der Pol\cite{van_der_Pol_1926}, Lienard\cite{Lienard1928} and Fatou\cite{Fatou1928} in the 1920s, along with Bogolyubov, Krylov and Mitropolskii\cite{Kryloff_1937,Bogoliubov1947,Bogoliubov1961,mitropolskiui1965} in the 1930s and 1940s, have shaped standard theories such as averaging and perturbation methods, \textbf{h}armonic \textbf{b}alance \textbf{m}ethod (HBM) and \textbf{d}escribing \textbf{f}unctions (DFs)\cite{Gelb1968}. 
In the 1950s, the \textbf{g}eneralized \textbf{f}requency \textbf{r}esponse \textbf{f}unctions (GFRFs)\cite{george1959} were proposed as a generalization of the FRFs. They are defined as the multidimensional Fourier transforms of the corresponding Volterra kernels in the time-domain series\cite{billing_JONES_1989}. 
While these multidimensional Fourier transforms can be difficult to measure, display and
interpret in practice, thus, a one-dimensional extension of the FRFs, known as the \textbf{n}onlinear \textbf{o}utput \textbf{f}requency \textbf{r}esponse \textbf{f}unction (NOFRF) method\cite{Lang__2005,Peng2007}, was proposed in the 2000s. The computation of the NOFRF is still under investigation, with recent methods based on \textbf{g}eneralized \textbf{a}ssociated \textbf{l}inear \textbf{e}quations (GALEs)\cite{Zhu_2022,Zhu2024}.

A wide variety of modifications to HBM have been developed over the last 40 years, including the \textbf{i}ncremental \textbf{h}armonic \textbf{b}alance (IHB)\cite{Lau_1982,cheunglau1982,Lau_1983,Pierre_1985,CHEUNG1990}, the \textbf{a}lternating \textbf{f}requency/\textbf{t}ime domain method using \textbf{h}armonic \textbf{b}alance (AFTHB)\cite{Cameron_1989}, the \textbf{a}daptive \textbf{h}armonic \textbf{b}alance (AHB)\cite{MAPLE2004,Lin2023}, and various further improvements and modifications of these methods\cite{Gilmore1986,Wang2015,RAHMAN2018893,Sharif_2019,Wu_2019,Ullah2020,Hosen_2020}. Comprehensive reviews of these methods can be found in Refs.~\cite{CHENG2017340,RIJLAARSDAM201711nuij,LIN2018270,LU2021116406,Zhu_2022}. While initially developed for system analysis, these methods have also found applications in system identification over the last 20 years (see, e.g., Refs.~\cite{Thothadri_2003,NOEL20172,Miguel2021,Tagipour2022}). In these approaches, the system structure (i.e., the degree or the nonlinearity or the explicit expression of the \textbf{n}onlinear \textbf{d}ifferential \textbf{e}quation (NDE) defining the system) must be predefined. This often entails solving numerous algebraic equations for each system. This limitation emphasizes the challenge of establishing a general system identification method applicable to all nonlinear systems. 



In this direction, the seeking of practical implementations for general system identification methods has led to the development of the \textbf{n}onlinear \textbf{a}uto\textbf{r}egressive \textbf{m}oving \textbf{a}verage model with e\textbf{x}ogenous inputs (NARMAX model)\cite{LEONTARITIS_1985,LEONTARITIS_1985_ii} in the 1980s. 
This model is versatile, applicable to a wide range of nonlinear systems, and has evolved into a philosophy of nonlinear system identification\cite{BILLINGS1989Tsang,Pearson1999,Billings_2013}. This philosophy aims to obtain a parsimonious model, which is a system with as few parameters and maximum lags as possible that can adequately describe the phenomena under study. 
Essentially, the NARMAX method considers the output as a nonlinear function of the previous values (input, output, and noise) up to specified time lags (also known as maximum lags). 
The model parameters are then estimated from input-output data using methods like \textbf{o}rthogonal \textbf{l}east \textbf{s}quares (OLS)\cite{Chen1989Billings_OLS} or \textbf{f}orward \textbf{r}egression OLS (FROLS)\cite{Billings1988_FROLS,Billings1989_FROLS}. This facilitates the fitting of the parsimonious model (referring here to the model with the fewest parameters and maximum lags from the given input-output data).  
However, practical application of the NARMAX method often results in various potentially suitable models based on the given input-output data. Choosing the most appropriate model becomes nontrivial without additional information about the system dynamics\cite{Sun2023}.  
Therefore, NARMAX technique is classified as a parametric black-box method (parametric because the system parameters are explicitly estimated through fitting, and black-box because no information about the system structure is required during the identification process).

During the last 15 years, data-driven algorithms have emerged for system identification of NDEs based on sparse regression and machine learning techniques\cite{Brunton_Kutz_2022}. A notable advancement is the \textbf{s}parse \textbf{i}dentification of \textbf{n}onlinear \textbf{dy}namics (SINDy)\cite{brunton2016}, which utilizes a sparsity-promoting framework to identify interpretable models from data by  selecting the most dominant candidate terms from a high-dimensional nonlinear function space. This technique has been improved, incorporating Bayesian sparse regression\cite{Hirsh2022} and ensemble method to estimate inclusion of probabilities\cite{Fasel2022}. Recently, the \textbf{a}utomatic \textbf{r}egression for \textbf{go}verning equation\textbf{s} (ARGOS)\cite{Egan2024} approach was developed to automatically address issues related to parameter selection for the Savitzky-Golay filter (to both reduce noise and compute numerical derivatives) and also the selection of the hyperparameters used for the identification of the interpretable model from the design matrix.   
Additionally, other related methodologies have been developed, such us the SINDy-\textbf{s}ensitivity \textbf{a}nalysis (SINDy-SA)\cite{Naozuka2022} framework, the model-LISS algorithms\cite{Strebel2023}, the \textbf{j}oint \textbf{m}aximum \textbf{a} \textbf{p}osteriori (JMAP) and \textbf{v}ariational \textbf{B}ayesian \textbf{a}pproximation (VBA)\cite{Niven2019}, and \textbf{s}tochastic \textbf{a}pproximation \textbf{M}onte \textbf{C}arlo (SAMC)\cite{LiuWangCao2018}.

These sparse regression algorithms are essentially based on a design matrix that contains a list of nonlinear functions. Using some minimization algorithm and threshold parameters, one or a linear combination of the functions from the list are found to describe the main terms of the nonlinear system \cite{brunton2016,Schaeffer2018}. 
However, two limitations of this approach are: (i) there is not guarantee that the method will perform well in identifying all systems, especially when dealing with complex nonlinearities that can not be well
approximated with the basis functions used in the design matrix\cite{Egan2024}; and (ii), the non unicity of the obtained model. 
To illustrate the latter issue, consider a simple example: suppose the design matrix contains both sinusoidal and polynomial functions, and the nonlinearity is given by f(x)=1-cos(x), where the variable x (which is an input data for the modeling technique) ranges from -$\pi$/4 to $\pi$/4. In this case, both models $\hat{f}_1$(x)$=$0.4819 x$^2$ and $\hat{f}_2$(x)=1-cos(x) are good representations of the system. In this example, adhering to the parsimonious principle\cite{Billings_2013,Kutz2022}, we should choose $\hat{f}_1$ due to its simpler expression. 
In summary, this non-uniqueness of the obtained model unavoidably arises when the basis functions used for constructing the design matrix are not orthogonal.

Conversely, the Fourier spectrum of a signal (i.e., Fourier coefficients obtained, e.g., from \textbf{f}ast \textbf{F}ourier \textbf{t}ransform (FFT) algorithm) is uniquely defined. This uniqueness has motivated the developments of the \textbf{o}utput \textbf{f}requency \textbf{r}esponse \textbf{f}unction (OFRF)\cite{lang1996billings,Lang1997billings,LANG2007805}, and the \textbf{h}igher-\textbf{o}rder \textbf{s}inusoidal \textbf{i}nput \textbf{d}escribing \textbf{f}unction (HOSIDF)\cite{2006Nuij} from the 1990s to the 2000s.     
An ultimate goal of frequency-domain methods for nonlinear system estimation is to provide a clear, explicit relationship between the system Fourier spectrum computed from input-output data and the parameters defining the nonlinear system\cite{CHENG2017340}. 
The OFRF and HOSIDF methods have significantly contributed to this objective.  
In particular, OFRF employs a recursive algorithm to obtain model parameters\cite{LANG2007805} from the Fourier spectrum, and HOSIDF introduces a \emph{virtual harmonic generator} to study the response of each harmonic separately. 
Both of these techniques fall into the category of parametric grey-box methods. Here, grey-box signifies that the system structure (in this case the explicit expression of the NDEs) must be predefined, and parametric implies that model parameters are explicitly obtained by some minimization algorithm. Thus, explicit expressions of the NARMAX model or NDEs must be specified, leading to the necessity of performing numerous algebraic operations for each particular system under study.

Motivated also by this uniqueness of the Fourier spectrum of a signal, a recent alternative method for system identification and simulation has been recently proposed based on \textbf{c}haracteristic \textbf{c}urves (CCs)\cite{Gonzalez2023}.  
In particular, two different parametrizations (power series- and Fourier analysis-based models) of the characteristic curves for a first-order nonlinear system have been demonstrated to be mathematically equivalent. 
Specifically, starting from the definition of the Fourier series, an explicit expression has been derived linking the Fourier coefficients of the driven force to the parameters that define the CCs of a first-order nonlinear system when the dynamical variable is a single tone.
Hence, the FFT coefficients of the driven force are uniquely related to the parameters of a power series defining the CCs. 
This method is referred to as \textbf{s}inusoidal \textbf{o}utput \textbf{r}esponse in \textbf{p}ower \textbf{s}eries (SORPS), which is more appropiately named than the previously used term of \textbf{s}inusoidal \textbf{i}nput \textbf{r}esponse in \textbf{p}ower \textbf{s}eries (SIRPS). This distinction becomes evident when rewriting the NDE as a NARMAX model (see Appendix~\ref{app:2}). 

Hence, SORPS method can be classified as a grey-box model based on CCs. Grey-box means that some information about the system must be previously stated. In this case, it is the supposition that the input-output data is related by a first-order NDE that follows certain functional dependence. Furthermore, the concept of CCs means that the importance is given to the evaluation of the characteristic curves as a whole and not to the individual parameters of the model.

The SORPS formalism introduced in Ref.~\cite{Gonzalez2023} exhibits certain limitations. Firstly, it is applicable only to first-order nonlinear systems or, alternatively, to two variables of a system related through a first-order NDE. Secondly, it requires setting a sinusoidal single tone for the dynamical variable and a fixed time step in order to obtain the system identification. However, in real-world systems, factors such as equipment or process limitations, or simply due to the presence of noise, may make it impossible to establish a sinusoidal single tone for the dynamical variable. Thirdly, due to the divergent behavior of the power series, some limitations or special considerations must be taken into account regarding the range of variation of the dynamical variable.

In this work, we extend the SORPS formalism in order to address most of these limitations. Specifically, instead of relying on Fourier analysis from FFT for the system identification, as employed in Ref.~\cite{Gonzalez2023}, we consider here a modeling technique based on a power series-based model and relying on least-squares regression. 
As a result, this approach enables us to obtain the characteristic curves directly from any input-output \textbf{d}ata\textbf{s}et (DS), overcoming the previous limitation of using a single tone for the dynamical variable and a fixed time step.  Therefore, the second limitation is successfully addressed in this work. 
Moreover, the mathematical expressions obtained here bring more evident the third limitation, which is briefly discussed. However, a detailed analysis to address this limitation requires careful consideration, which is beyond the scope of this work.  
Although the first limitation is not addressed here, the results of this work   
represents a significant advancement in order to extend the SORPS formalism to higher-order nonlinear systems based on the use of CCs.   

While the SORPS technique can be expressed as a NARMAX model, it conceptually differs from the NARMAX philosophy (as well as sparse regression methodologies\cite{Kutz2022}), because, instead of emphasizing the fitting of a nonlinear model with as few parameters as possible, or finding the leading terms of the system, SORPS focuses on identifying the CCs that define the NDE. 
Although the SORPS method cannot be directly applied when the input-output variables are related by an NDE with an order higher than one, it can still be utilized by decomposing the system into a set of variables related by a functional dependence given by a first-order NDE (see Appendix~\ref{app:4}). 

Additionally, an alternative approach involves defining higher-order NDEs based on the concept of CCs. This latter represents a promising direction for future developments (see Sec.~\ref{sec:discussion}). 
The advantages of a CCs-based approach are twofold, the first one is that the system is uniquely defined when the functional structure is specified, therefore the identified system is uniquely defined for a given input-output data, overcoming the non-uniqueness issues of sparse regression methods. The second one is that the use of a predefined functional structure allows the computation of all the properties and algebraic manipulations beforehand, therefore, the same method can be applied for all the systems that are defined with the given functional dependence, without any additional consideration for the particular system under study, overcoming many practical challenges associated with the implementation of the NOFRF, OFRF, HOSIDF and HBM-based methods.

Hence, these CCs-based methods offer an additional (or complementary) perspective compared to NARMAX-philosophy and NDEs-based methods, such as sparse regression techniques. This new perspective involves defining a NDE based on CCs for every higher-order system (this work focusing only on a first-order system). 
Subsequently, the parsimonious model can be determined by identifying the lowest-order system from the list of NDEs that most accurately reproduces the input-output data. 
Further discussion on how the concept of CCs could potentially be applied to higher-order systems is addressed in Sec.~\ref{sec:discussion}.

The work is structured as follows. 
The Formalism section is composed by: Sec.~\ref{sec:formalism:modeling} (model 1) presents the power series expression used for system modeling, with hyperparameters $\hat{A}_0$ and $\hat{A}_1$, and whose model parameters are obtained by the least-squares regression method as detailed in Appendix~\ref{app:1}. 
Sec.~\ref{sec:formalism:resultA} (model 2) presents the connection to another power series-based model with hyperparameters $\hat{A}_0$ and $\hat{A}_1$, which is useful for comparing the model parameters obtained from different DSs;  
Sec.~\ref{sec:formalism:resultB} (model 3) presents the connection between the parameters of the power series-based model to those corresponding to the Fourier series-based model using the SORPS formalism; 
Sec.~\ref{sec:formalism:procedure} discusses the procedure for system identification and the steps to follow for transforming between the models; 
and Sec.~\ref{sec:formalism:considerations} discusses some considerations related to the formalism in relation to the issues of divergence and extrapolation.  
Also, an Illustrative example is discussed in Sec.~\ref{sec:illustrative}, where we compare the parameters of models 2 and 3 for different DSs and also study the system modeling in the presence of noise. 
Section~\ref{sec:discussion} discusses the limitations and future research directions, while conclusions are presented in Sec.~\ref{sec:conclusion}.
Appendixes~\ref{app:2},\ref{app:3}, and \ref{app:4} address the main points of the SORPS formalism, an alternative matrix formulation and limitations of the SORPS formalism, respectively.

\section{Formalism}\label{sec:formalism}
In circuit theory, signal processing and mechanical systems, it is useful to express a first-order nonlinear system as the following autonomous nonlinear system\cite{Oppenheim1999,Stoica2005,NumRecipesPress2007,Proakis20074th}

\begin{equation}
    y(t)= f(x(t))+g(x(t))x'(t) \, ,
    \label{eq:1ordsys}
\end{equation}
where $x(t)$ is the dynamical variable, $y(t)$ is the driven force, and $f(x)$ and $g(x)$ are commonly referred to as characteristic curves (CCs) since they define the system\cite{Gonzalez2023}. If these curves are known based on some parametrization, then the system given by Eq.~(\ref{eq:1ordsys}) is completely defined. 

In sections~\ref{sec:formalism:modeling}, \ref{sec:formalism:resultA} and \ref{sec:formalism:resultB}, we present three models consisting of different parametrizations for the characteristic curves and demonstrate that they are mathematically equivalent. Hence, by using these mathematical expressions, all models can be derived from knowledge of only one. Thus, model 1 will be used for system estimation, and model 2 and 3 will be obtained from model 1 by using the aforementioned mathematical expressions. Although these three models predict identical output results for a given input data, their individual parameters differ. Thus, models 2 and 3 are used for comparing the individual parameters obtained from different DSs. 

\subsection{Model 1: power series-based model with hyperparameters $\hat{A}_0$ and $\hat{A}_1$}\label{sec:formalism:modeling}
Assume that the characteristic curves of Eq.~\ref{eq:1ordsys} can be parameterized as a power series

\begin{equation}
\begin{aligned}
    f(x(t)) &=\sum_{j=0}^{N} \hat{f}_j \,\left(\frac{x(t)-\hat{A}_0}{\hat{A}_1}\right)^j  \\
    g(x(t)) &=\sum_{j=0}^{N-1} \hat{g}_j \, \left(\frac{x(t)-\hat{A}_0}{\hat{A}_1}\right)^j   \; , 
    \label{eq:1ordsys:fgtilde}
\end{aligned}
\end{equation}

where $N$ is the maximum desired order for the polynomial expansion, and $\hat{A}_0\in \mathbb{R}$ and $\hat{A}_1>0$ are \emph{hyperparameters} (tuning parameters) that have to be determined with some criterion. 
In this work, we utilize Eq.~\ref{eq:1ordsys:fgtilde} for system modeling. 
The modeling process involves determining the parameters $\hat{f}_j$ and $\hat{g}_j$ from a provided input-output \textbf{d}ata\textbf{s}et (DS) through the least-squares regression method, as explained in detailed in Appendix~\ref{app:1}. 
A DS is defined by the values of $x(t)$ and their corresponding values of $y(t)$ for all the values of $t$ in the interval $[0,T_s]$. 
By recognizing that the power series in Eq.~\ref{eq:1ordsys:fgtilde} diverge for $\left|(x(t)-\hat{A}_0)/\hat{A}_1\right| > 1$, we define the hyperparameters as

\begin{equation}
    \begin{aligned}
        \hat{A}_0 &\coloneqq \frac{\text{max}[x(t)]+\text{min}[x(t)]}{2} \\
        \hat{A}_1 &\coloneqq \frac{\text{max}[x(t)]-\text{min}[x(t)]}{2} \, ,
    \end{aligned}
    \label{eq:1ordsys:A0A1def}
\end{equation}

where $\text{min}[x(t)]$ and $\text{max}[x(t)]$ denote the minimum and maximum values of $x(t)$ over a specified time interval $t=[0,T_s]$, and $T_s$ represents, for instance, the total simulation time or the total duration of an experimental measurement. 
This definition guarantees that $\left|(x(t)-\hat{A}_0)/\hat{A}_1\right| \leq 1$ holds for all values of $t$ within the interval. 
The suitability of this definition is further discussed in Sec.~\ref{sec:illustrative:A0hatA1hat} using an illustrative example. For future reference, we express Eq.~\ref{eq:1ordsys:fgtilde} as a truncated Maclaurin series, obtaining

\begin{equation}
\begin{aligned}
    f(x(t)) &=\sum_{j=0}^{N} (x(t))^j \, \left[ \sum\limits_{k=j}^N  \binom{k}{j} \frac{\left( -\hat{A}_0\right)^{k-j}}{\hat{A}_1^k} \hat{f}_k\right] \\
    g(x(t)) &=\sum_{j=0}^{N-1} (x(t))^j \, \left[ \sum\limits_{k=j}^{N-1}  \binom{k}{j} \frac{\left( -\hat{A}_0\right)^{k-j}}{\hat{A}_1^k} \hat{g}_k\right]   \; , 
    \label{eq:1ordsys:fgtilde:pow}
\end{aligned}
\end{equation}

For completeness, we present the intermediate steps for the demonstration of Eq.~\ref{eq:1ordsys:fgtilde:pow} in the following

\begin{equation}
\begin{aligned}
    f(x(t))&=\sum_{k=0}^{N} \hat{f}_k \,\left(\frac{x(t)-\hat{A}_0}{\hat{A}_1}\right)^k \\ &= \sum_{k=0}^{N} \hat{f}_k\, \left[\sum_{j=0}^k \binom{k}{j} (x(t))^{j} \frac{\left( -\hat{A}_0\right)^{k-j}}{\hat{A}_1^k} \right] \\ &=  \sum_{j=0}^{N} (x(t))^j \, \left[ \sum\limits_{k=j}^N  \binom{k}{j} \frac{\left( -\hat{A}_0\right)^{k-j}}{\hat{A}_1^k} \hat{f}_k\right] \; ,
    \label{eq:1ordsys:fexpanded}
\end{aligned}
\end{equation}

where the second equality employs the binomial theorem, and the third equality results from rearranging the order of summation. The demonstration of $g(x(t))$ is analogous.

\subsection{Model 2: power series-based model with hyperparameters $A_0$ and $A_1$}\label{sec:formalism:resultA}

Assume that the characteristic curves of Eq.~\ref{eq:1ordsys} can be parameterized as a power series

\begin{equation}
\begin{aligned}
    f(x(t)) &=\sum_{j=0}^{N} f_{\!j} \,\left(\frac{x(t)-A_0}{A_1}\right)^j  \\
    g(x(t)) &=\sum_{j=0}^{N-1} g_{\!j} \, \left(\frac{x(t)-A_0}{A_1}\right)^j   \; , 
    \label{eq:1ordsys:fgpol}
\end{aligned}
\end{equation}

where the hyperparameters $A_0\in \mathbb{R}$ and $A_1>0$ must be chosen with some criterion. A discussion regarding the selection of $A_0$ and $A_1$ is provided in Sec.~\ref{sec:formalism:considerations}.  
We aim to establish a link between parameters $\hat{f}_j$ and $\hat{g}_j$, obtained from the system modeling (Sec.~\ref{sec:formalism:modeling}), and parameters $f_{\!j}$ and $g_{\!j}$ defined in Eq.~\ref{eq:1ordsys:fgpol}.
To achieve this, we expand Eq.~\ref{eq:1ordsys:fgpol} using a procedure similar to Eq.~\ref{eq:1ordsys:fexpanded}, obtaining for $f(x)$

\begin{equation}
\begin{aligned}
    f(x(t))&=\sum_{k=0}^{N} f_k \,\left(\frac{x(t)-A_0}{A_1}\right)^k \\ &=  \sum_{j=0}^{N} (x(t))^j \, \left[ \sum\limits_{k=j}^N  \binom{k}{j} \frac{\left( -A_0\right)^{k-j}}{A_1^k} f_k\right]
    \label{eq:1ordsys:fexpanded:result}
\end{aligned}
\end{equation}

By comparing Eqs.~\ref{eq:1ordsys:fexpanded} and \ref{eq:1ordsys:fexpanded:result} we can match the coefficients corresponding to each power of $x(t)$, obtaining

\begin{equation}
\begin{aligned}
 \frac{f_{\!j}}{A_1^j} +\sum\limits_{k=j+1}^N  \binom{k}{j} \frac{\left( -A_0\right)^{k-j}}{A_1^k} f_k  = 
\frac{\hat{f}_j}{\hat{A}_1^j}+ \sum\limits_{k=j+1}^N \binom{k}{j} \frac{\left( -\hat{A}_0\right)^{k-j}}{\hat{A}_1^k} \hat{f}_k  \; \; ,
    \label{eq:1ordsys:fexpanded:igualation:demost}
\end{aligned}
\end{equation}

where the first term $k=j$ has been separated by convenience. Eq.~\ref{eq:1ordsys:fexpanded:igualation:demost}  can be solved for $f_{\!j}$ through back substitution as  (the equation for $g_{\!j}$ is the same but replacing $f$ by $g$ and $N$ by $N-1$)

\begin{equation}
\begin{aligned}
\begin{cases}
\displaystyle f_{\!j} = A_1^j \cdot \left\{
\frac{\hat{f}_j}{\hat{A}_1^j}+ \sum\limits_{k=j+1}^N \binom{k}{j} \left[ \frac{\left( -\hat{A}_0\right)^{k-j}}{\hat{A}_1^k} \hat{f}_k -\frac{\left( -A_0\right)^{k-j}}{A_1^k} f_k  \right] \right\}  &  \text{ for } j=\{N,N-1,\cdots,0\} \\ 
\displaystyle g_{\!j} = A_1^j \cdot \left\{
\frac{\hat{g}_j}{\hat{A}_1^j}+ \sum\limits_{k=j+1}^{N-1} \binom{k}{j} \left[ \frac{\left( -\hat{A}_0\right)^{k-j}}{\hat{A}_1^k} \hat{g}_k -\frac{\left( -A_0\right)^{k-j}}{A_1^k} g_k  \right] \right\}  & \text{ for }  j=\{N-1,N-2,\cdots,0\} \; ,
    \label{eq:1ordsys:fgexpanded:igualation}
\end{cases}
\end{aligned}
\end{equation}

where parameters $f_{\!j}$ and $g_{\!j}$ must be calculated in a decreasing order for $j$. 
Equation~\ref{eq:1ordsys:fgexpanded:igualation} allows us to obtain $f_{\!j}$ and $g_{\!j}$ from $\hat{f}_j$ and $\hat{g}_j$, this is useful for comparing parameters obtained from different DSs, as explained in Sec.~\ref{sec:formalism:procedure}.

Notice that Eq.~\ref{eq:1ordsys:fgexpanded:igualation} can be inverted by interchanging the variables with the \emph{hat} symbol $\widehat{\left.(\cdot)\right.}$ for the corresponding variables without the \emph{hat} symbol, obtaining the values $\hat{f}_j$ and $\hat{g}_j$ from $f_{\!j}$ and $g_{\!j}$.
Therefore, there is a one-to-one correspondence between parameters of the two power series. This correspondence implies that the two power series are simply two representations of the same model.
Equation~\ref{eq:1ordsys:fgexpanded:igualation} can also be written in a matrix formulation as discussed in Appendix~\ref{app:3}.

\subsection{Model 3: Fourier analysis-based model}\label{sec:formalism:resultB}

Consider the function $y(t)$ for $t$ defined in the interval of time $t\in[0,T]$, where $T=2\pi/\omega$. We can extend this function to the real domain by assuming that it repeats periodically with period $T$. Then, we can express its Fourier series as

\begin{equation}
    y(t)=a_0+\sum_{k=1}^\infty ( a_k \cos(k\omega t) + b_k \sin(k\omega t) ) \, ,
    \label{eq:formalism:fourierv}
\end{equation}

where $\{a_j\}$ and $\{b_j\}$ are the Fourier coefficients. In Ref. \cite{Gonzalez2023} (see Appendix~\ref{app:2} for more details), a connection has been found between the Fourier coefficients of $y(t)$ for a dynamical variable $x(t)=A_0+A_1 \sin(\omega t)$ and the parameters of a power series for the characteristic curves $f(x)$ and $g(x)$ for the system defined in Eq.~\ref{eq:1ordsys}. 
Specifically, the Fourier coefficients of $y(t)$ up to a given maximum order $N$ when the dynamical variable is a single tone $x(t)=A_0+A_1 \sin{\omega t}$ (i.e., the values of $\{a_j\}$ and $\{b_j\}$ with $j\in[0,N]$), are related to the parameters of a power series defined by Eq.~\ref{eq:1ordsys:fgpol} (i.e., the values of $f_{\!j}$ and $g_{\!j}$ with $j\in[0,N]$).  
This connection can be expressed as


\begin{equation}
\begin{aligned}
    f_{\!j}&=\sum_{\substack{k=j\\ k\leftarrow k+2}}^N [ab]_{kj} \left[ \sum_{\substack{l=k-j\\ l\leftarrow l+2 }}^k \binom{k}{l} \binom{l/2}{(l-k+j)/2} \right] \\ 
g_{\!j}&= A_1 \omega \sum_{\substack{k=j+1\\ k\leftarrow k+2}}^{N} [ab]_{kj} \left[ \sum_{\substack{l=k-j\\ l\leftarrow l+2 }}^k \binom{k}{l} \binom{(l-1)/2}{(l-k+j)/2} \right] \\ &= A_1 \omega \sum_{\substack{k=j\\ k\leftarrow k+2}}^{N-1} [ab]_{(k+1)j} \left[ \sum_{\substack{l=k-j\\ l\leftarrow l+2 }}^k \binom{k+1}{l+1} \binom{l/2}{(l-k+j)/2} \right] \; ,
\label{eq:fgj_abkj}
\end{aligned}
\end{equation}

where
\begin{align}
    [ab]_{kj}&\coloneqq \frac{1+(-1)^j}{2}(-1)^{\nicefrac{j}{2}}a_k+\frac{1+(-1)^{j-1}}{2}(-1)^{\nicefrac{(j-1)}{2}}b_k \nonumber \\&= \begin{cases}
    a_k(-1)^{\nicefrac{j}{2}} & \text{ if } j \text{ is even} \\ 
    b_k(-1)^{\nicefrac{(j-1)}{2}} & \text{ if } j \text{ is odd} \; .
    \end{cases}
    \label{eq:abkj}
\end{align}

See Appendix~\ref{app:2} for further details of this result. Equations~\ref{eq:fgj_abkj} and \ref{eq:abkj} can be compactly represented in vector form as follows 

\begin{equation}
\begin{aligned}
    \vec{f} &= \textbf{M}_{ab2f} \cdot \overrightarrow{\smash{ab}\vphantom{b}}_{\!f} \\
    \vec{g} &= A_1 \omega \; \textbf{M}_{ab2g} \cdot \overrightarrow{\smash{ab}\vphantom{b}}_{\!g}  \; ,
    \label{eq:sirps:vecab2fg}
\end{aligned}
\end{equation}

where

\begin{equation}
    \begin{aligned}
        \vec{f} &\coloneqq [f_0,f_1,\cdots,f_N]^T \\
        \vec{g} &\coloneqq [g_0,g_1,\cdots,g_{N-1}]^T \\
         \overrightarrow{\smash{ab}\vphantom{b}}_{\!f} &\coloneqq  \begin{cases}
            [a_0,b_1,a_2,b_3,\cdots,b_{N-1},a_N]^T & \text{ if } N \text{ is even} \\
            [a_0,b_1,a_2,b_3,\cdots,b_{N-1}]^T & \text{ if } N \text{ is odd}
        \end{cases}  \\
          \overrightarrow{\smash{ab}\vphantom{b}}_{\!g} &\coloneqq  \begin{cases}
            [a_1,b_2,a_3,b_4,\cdots,a_{N-1},b_N]^T & \text{ if } N \text{ is even} \\
            [a_1,b_2,a_3,b_4,\cdots,a_{N-1}]^T & \text{ if } N \text{ is odd ,}
        \end{cases}
        \label{eq:sirps:defs}
    \end{aligned}
\end{equation}

and the elements of the matrices $\textbf{M}_{ab2f}$ and $\textbf{M}_{ab2g}$ are defined by

\begin{equation}
(\textbf{M}_{ab2f})_{jk} \coloneqq
\begin{cases}
  \displaystyle(-1)^{\floor*{j/2}}  \sum_{\substack{l=k-j\\ l\leftarrow l+2 }}^k \binom{k}{l} \binom{l/2}{(l-k+j)/2}   & \text{ if } j,k \text{ are both even or both odd and } k \geq j \\
    0 & \text{ other case } 
\end{cases}
\label{eq:sirps:ab2f}
\end{equation}

and

\begin{equation}
(\textbf{M}_{ab2g})_{jk} \coloneqq
\begin{cases}
  \displaystyle(-1)^{\floor*{j/2}}  \sum_{\substack{l=k-j\\ l\leftarrow l+2 }}^k \binom{k+1}{l+1} \binom{l/2}{(l-k+j)/2}   & \text{ if } j,k \text{ are both even or both odd and } k \geq j \\
    0 & \text{ other case   ,} 
\end{cases}
\label{eq:sirps:ab2g}
\end{equation}
where $j$ and $k$ represent the row and column indexes, respectively. Also, $\lfloor \cdot \rfloor$ represents the floor function, and the indexing convention starts from zero. Specifically, for $\textbf{M}_{ab2f}$, indexes range from $0$ to $N$, and for $\textbf{M}_{ab2g}$, from $0$ to $N-1$.
$\textbf{M}_{ab2f}$ and $\textbf{M}_{ab2g}$ are upper triangular matrices, and can be solved by back substitution.

In summary, Eq.~\ref{eq:sirps:vecab2fg} with the definitions given in Eqs.~\ref{eq:sirps:defs}, \ref{eq:sirps:ab2f} and \ref{eq:sirps:ab2g} allows us to obtain the parameters $f_{\!j}$ and $g_{\!j}$ defined in Eq.~\ref{eq:1ordsys:fgpol} based on the Fourier coefficients defined in Eq.~\ref{eq:formalism:fourierv}, where the Fourier coefficients correspond to the Fourier series of the driven force $y(t)$ when the dynamical variable is a single tone $x(t)=A_0+A_1 \sin(\omega t)$. 
Complementary, Eq.~\ref{eq:sirps:vecab2fg} also allows us to obtain  $\overrightarrow{\smash{ab}\vphantom{b}}_{\!f}$ and $\overrightarrow{\smash{ab}\vphantom{b}}_{\!g}$ from $\vec{f}$ and $\vec{g}$. This correspond to the process for obtaining the Fourier coefficients of $y(t)$ for a \emph{hypothetical} single tone in $x(t)$ based on parameters $f_{\!j}$ and $g_{\!j}$. 
We will use this result to obtain the parameters of the Fourier analysis-based model from the  power series-based parameters, obtained from model 1 from arbitrary input-output data for a given values of $A_0$, $A_1$ and $\omega$.

\subsection{Procedure for system identification}\label{sec:formalism:procedure}

We use three models in this work. The first one is for system identification purposes, and second and third ones are alternative parametrizations facilitating the comparison between model parameters obtained from different DSs. 
DSs can be generated from experimental data or simulations. For the latter case, two alternatives are possible: the first one by setting $x(t)$ and obtaining the corresponding values of $y(t)$ by evaluating the first-order system of Eq~\ref{eq:1ordsys}; and the second one by setting $y(t)$ and obtaining the corresponding values of $x(t)$ by numerically integrating the first-order system.

\begin{itemize}
   \item[$\blacksquare$] Model 1 (power series with hyperparameters $\hat{A}_0$ and $\hat{A}_1$): Based on a given DS,  
    calculate the model parameters $\hat{f}_j$ and $\hat{g}_j$ for the power series defined in Eq.~\ref{eq:1ordsys:fgtilde} using the least-squares regression method. Utilize the definitions of $\hat{A}_0$ and $\hat{A}_1$ given in Eq.~\ref{eq:1ordsys:A0A1def}.
   \item[$\blacksquare$] Model 2 (power series with hyperparameters $A_0$ and $A_1$):  Although model 1 is enough for practical system identification purposes, it happens that $\hat{A}_0$ and $\hat{A}_1$ depend on the range of values of the dynamical variable, thus, parameters $\hat{f}_j$ and $\hat{g}_j$ obtained from different DSs can not be compared between them.   
To facilitate this comparison, we propose to use model 2, which is mathematically equivalent to model 1 but with other hyperparameters $A_0$ and $A_1$ that are maintained fixed for all the DSs used for system estimation. Model 2 is obtained as follows: firstly, obtain model parameters $\hat{f}_j$ and $\hat{g}_j$ from model 1; secondly; use Eq.~\ref{eq:1ordsys:fgexpanded:igualation} (or the matrix formulation of Eq.~\ref{eq:1ordsys:fghat2fg} from Appendix~\ref{app:3}) to obtain parameters $f_{\!j}$ and $g_{\!j}$ from $\hat{f}_j$ and $\hat{g}_j$. For the second step, define the values for hyperparameters $A_0$ and $A_1$ and maintain those fixed for all the DSs used for system estimation. Consequently, model 2 allows us to directly compare parameters $f_{\!j}$ and $g_{\!j}$ obtained from different DSs.  
   \item[$\blacksquare$] Model 3 (Fourier analysis-based model): Using the SORPS formalism, we can use the parameters from model 2 to calculate the model parameters of the Fourier analysis-based model. Hence, we can compare parameters of Fourier analysis-based model that were obtained from different DSs. To achieve this, use Eq.~\ref{eq:sirps:vecab2fg} to obtain Fourier coefficients $a_j$ and $b_j$ of the Fourier series of $y(t)$ (or equivalently, the elements of the vectors $\overrightarrow{\smash{ab}\vphantom{b}}_{\!f}$ and $\overrightarrow{\smash{ab}\vphantom{b}}_{\!g}$ defined in Eq.~\ref{eq:formalism:fourierv})  from parameters $f_{\!j}$ and $g_{\!j}$ of model 2. 
Define the value of $\omega$ and use the same values of $A_0$ and $A_1$ defined in Step 2. This procedure can be written in vector form as

\begin{equation}
    \begin{aligned}
     \overrightarrow{\smash{ab}\vphantom{b}}_{\!f}  &= \textbf{M}_{ab2f}^{-1}\cdot  \vec{f} \\    
      \overrightarrow{\smash{ab}\vphantom{b}}_{\!g} &= \frac{1}{A_1 \omega} \textbf{M}_{ab2g}^{-1}  \cdot \vec{g} .      
    \end{aligned}  
    \label{eq:fg2abfg}
\end{equation}
\end{itemize}
 
Notice that Eq.~\ref{eq:1ordsys:fgexpanded:igualation} establishes one-to-one correspondences between $\hat{\vec{f}}$ and $\vec{f}$, as well as between $\hat{\vec{g}}$ and $\vec{g}$, i.e., we can transform from \{$\hat{\vec{f}}$,$\hat{\vec{g}}$\} to \{$\vec{f}$,$\vec{g}$\}, and vice versa. Consequently, if we evaluate the characteristic curves $f(x)$ and $g(x)$ from model 1 or model 2, we obtain the same values of $y(t)$. Essentially, 
we are representing the same system using another parametrization, and the predicted values must coincide.  The same applies to model 3 (Eq.~\ref{eq:fgj_abkj}). 
Therefore, the three models of this work yield the same predicted values of $y(t)$. This result is not surprising, given they are different parametrizations of the same system which are mathematically equivalent. 
However, notice that the parameters obtained for models 1 and 2 (the power series-based models with hyperparameters $\hat{A}_0$ and $\hat{A}_1$, and with $A_0$ and $A_1$, respectively) are different when $\hat{A}_0 \neq A_0$ or $\hat{A}_1 \neq A_1$. Furthermore, parameters of model 3 are different than those of models 1 and 2. 
Therefore, the individual parameters obtained for these three models are different. 
This motivates a deeper study of the properties of the different model parameters, which is addressed in this work through the illustrative example of Sec.~\ref{sec:illustrative}.

\subsection{Considerations related to extrapolation issues}\label{sec:formalism:considerations}

In this section, we discuss some issues related to the divergent behavior of the power series and the extrapolation of the model to ranges of $x$ beyond the interval used for system modeling. 
Given the mathematical equality in the predicted values $\hat{y}(t)$ between the three models, this analysis of divergence can be conducted through any of the three models. 
We will consider the first model, defined by the power series with hyperparameters $\hat{A}_0$ and $\hat{A}_1$. Specifically, consider the characteristic curve for $f(x)$ given in Eq.~\ref{eq:1ordsys:fgtilde}, rewritten in the following for convenience

\begin{equation}
\begin{aligned}
    f(x(t)) &=\sum_{j=0}^{N} \hat{f}_{\!j} \,\left(\frac{x(t)-\hat{A}_0}{\hat{A}_1}\right)^j     \; . 
    \label{eq:1ordsys:fgtilde:illustrative}
\end{aligned}
\end{equation}

Suppose $\hat{A}_0=0$ and $\hat{A}_1=1$ to simplify the analysis. 
If $f_{\!j} = 0$ for all $j>0$, then $f(x)=f_{\!0}$ is a constant.
Otherwise, suppose there is at least one index $j>0$ where $\hat{f}_{\!j}\neq 0$. Define $M$ as the greater value of $j$ such as $\hat{f}_{\!j}\neq 0$ (thus, $\hat{f}_{\!M}\neq 0$). Then, the term with the highest power ($\hat{f}_{\!M} x^M$) dominates the behavior of the polynomial.

If $x$ tends to infinity ($x\rightarrow +\infty$), the function $f(x)$ tends to $+\infty$ or $-\infty$, depending if $\hat{f}_{\!M}>0$ or $\hat{f}_{\!M}<0$, respectively.
Similarly, when $x\rightarrow-\infty$, the function $f(x)$ tends to  $+\infty$ for $\hat{f}_{\!M}>0$ and $M$ even, or $\hat{f}_{\!M}<0$ and $M$ odd. On the contrary, $f(x)$ tends to $-\infty$ for $\hat{f}_{\!M}>0$ and $M$ odd, or $\hat{f}_{\!M}<0$ and $M$ even.
For other hyperparameter values for $\hat{A}_0$ and $\hat{A}_1$, i.e. $\hat{A}_0\neq 0$ and $\hat{A}_1\neq 1$, the analysis is analogous.

To summarize, if $\{f(x); -\infty<x <\infty\}$ is a polynomial whose degree is non-zero, then $|f(x)|$ becomes unbounded as $|x|$ tends to infinity (see Ref.~\cite{Powell1981} for more details). 
This has as a consequence that power series can not represent functions $f(x)$ whose limits values are a constant up to arbitrary range of values of $x$ (with the exception of the constant case). 
Consequently, supposing that the function $f(x)$ has a constant as limit value for $x\rightarrow \infty$ or $x\rightarrow -\infty$, and that, we have the information of $f(x)$ for a given interval of $x$, then, the power series obtained from the fit is expected to represent correctly the behavior of $f(x)$ inside the interval. But if we evaluate the power series obtained from the fit for values of $x$ beyond the interval, the predicted values from the power series will separate increasingly as we move away of the interval, due to the mentioned divergent property of the power series. 

When we evaluate the power series obtained from the fit for values of $x$ beyond the interval that is used for the fit, we are in fact doing an extrapolation of the model. 
Here, we have demonstrated that if the function $f(x)$ has a constant as limit value for $x\rightarrow \infty$ or $x\rightarrow -\infty$, then the extrapolation will fail increasingly as we move away of the interval. 
For example, if $f(x)=exp(-x^2)$ and $x$ is defined on a small interval [a,b], the obtained model will be expected to be good inside this interval, but it will fail beyond that.

On the contrary, suppose the case where the function $f(x)$ has divergent behaviors for both  $x\rightarrow \infty$ and $x\rightarrow -\infty$. In this case, it is not obvious if the extrapolation of the model will fail as we move away of the interval. For example, a polynomial fit of a $x^2$ function, where $x$ ranges from -1 to 1, is enough to properly identify the function in the whole domain from $-\infty$ to $\infty$. 
Therefore, if $f(x)$ is a polynomial, it is possible that the fitting using a power series with $x$ defined only on a small interval, will be enough to determine the polynomial coefficients that are valid for the whole domain. In this case, the extrapolation of the model for values of $x$ beyond the interval used for the fit is expected to be good.

Further studies about extrapolation issues require a careful analysis, that is beyond of the scope of this work, 
as our focus is on comparing parameters obtained for different DSs as they become more complex. 
The \emph{complexity} of the DS, in this context, can be quantified, for example, by the condition number of the least-squares regression matrix for the system estimation (defined in Appendix~\ref{app:1}). Thus, the condition number increases as the DS becomes more complex. 
Therefore, to ensure that only those errors arising from the complexity of different input-output DSs, and also, to assess a fair comparison between different DSs, we impose the restriction that 
all the $x(t)$ functions across the DSs have a range greater than the
interval $[A_0-A_1,A_0+A_1]$. It can be mathematically expressed as $[A_0-A_1,A_0+A_1]\subseteq [\hat{A}_0-\hat{A}_1,\hat{A}_0+\hat{A}_1]=[\text{min}[x(t)],\text{max}[x(t)]]$, for all $x(t)$ across the different DSs, where Eq.~\ref{eq:1ordsys:A0A1def} is used in the last equality. 
With this choice, and by assuring us that all simulations with the identified models satisfy $x\in[A_0-A_1,A_0+A_1]$, we guarantee that only those errors arising from the complexity of different DSs are present in the simulations.

\section{Illustrative example}
\label{sec:illustrative}
In this section, we utilize an illustrative example to demonstrate the application of the formalism to a given system, highlighting the advantages and disadvantages of the formulation presented in this work. 
To begin, we define the characteristic curves of the system as

\begin{equation}
    \begin{aligned}
        f(x(t)) &= \sum_{i=0}^{6} f^{th}_i x(t)^i \\
        g(x(t)) &=1+e^{-x(t)^2} \; ,
    \end{aligned}
    \label{eq:example:charcuves}
\end{equation}
where the polynomial coefficients are defined as follows: $f^{th}_0$=~1, $f^{th}_1$=~-1, $f^{th}_2$=~0.5, $f^{th}_3$=~3.9, $f^{th}_4$=~-3, $f^{th}_5$=~-0.2, and $f^{th}_6$=~0.4, with the superscript $^{th}$ denoting that these are theoretical values.

\subsection{Considerations about the definitions of $\hat{A}_0$ and $\hat{A}_1$}\label{sec:illustrative:A0hatA1hat}
In this section, we evaluate the suitability of the definitions provided in Eq.~\ref{eq:1ordsys:A0A1def} for the hyperparameters $\hat{A}_0$ and $\hat{A}_1$. 
For this analysis, we use for system modeling a DS generated by setting $x(t)=0.5+1.5\sin(\omega_s t)$ and calculating $y(t)$ from Eq.~\ref{eq:1ordsys}, through the evaluation of the characteristic curves defined in Eq.~\ref{eq:example:charcuves}. Based on the proposed definitions for $\hat{A}_0$ and $\hat{A}_1$, we obtain $\hat{A}_0=0.5$ and $\hat{A}_1=1.5$ for this DS.  
Here, the frequency is $\omega_s=2\pi/T_s$, and the period $T_s$ is set equal to the total simulation time.   
The total simulation time is $T_s=100$ s, with a simulation step time of $\Delta t=0.01$ s. Consequently, the vectors $\vec{x}$ and $\vec{y}$ have a length $L=1001$, where the vector $\vec{t}$ starts at $t=0$ and ends with $t=100$ s.

Firstly, we obtain model 1 by calculating the model parameters $\hat{f}_j$ and $\hat{g}_j$ in Eq.~\ref{eq:1ordsys:fgtilde} through a least-squares regression, as detailed in Appendix~\ref{app:1}, from the given DS. We define the maximum order $N=20$ for the polynomial expansion (a discussion about the selection of the $N$ value is given in Sec.~\ref{sec:illustrative:maxN}).
To facilitate the analysis, we introduce alternative labels called as $\tilde{A}_0$ and $\tilde{A}_1$, for the parameter definitions, where $\tilde{A}_1$ will be varied from $0.01\hat{A}_1$ to  $1000\hat{A}_1$, and $\tilde{A}_0$ from $\hat{A}_0-4$ to $\hat{A}_0+4$. Hence, we obtain system modeling using model 1 by using different values of $\tilde{A}_0$ and $\tilde{A}_1$ in the mentioned range of values.

\begin{figure}[!ht]
    \centering
    \includegraphics[scale=0.4]{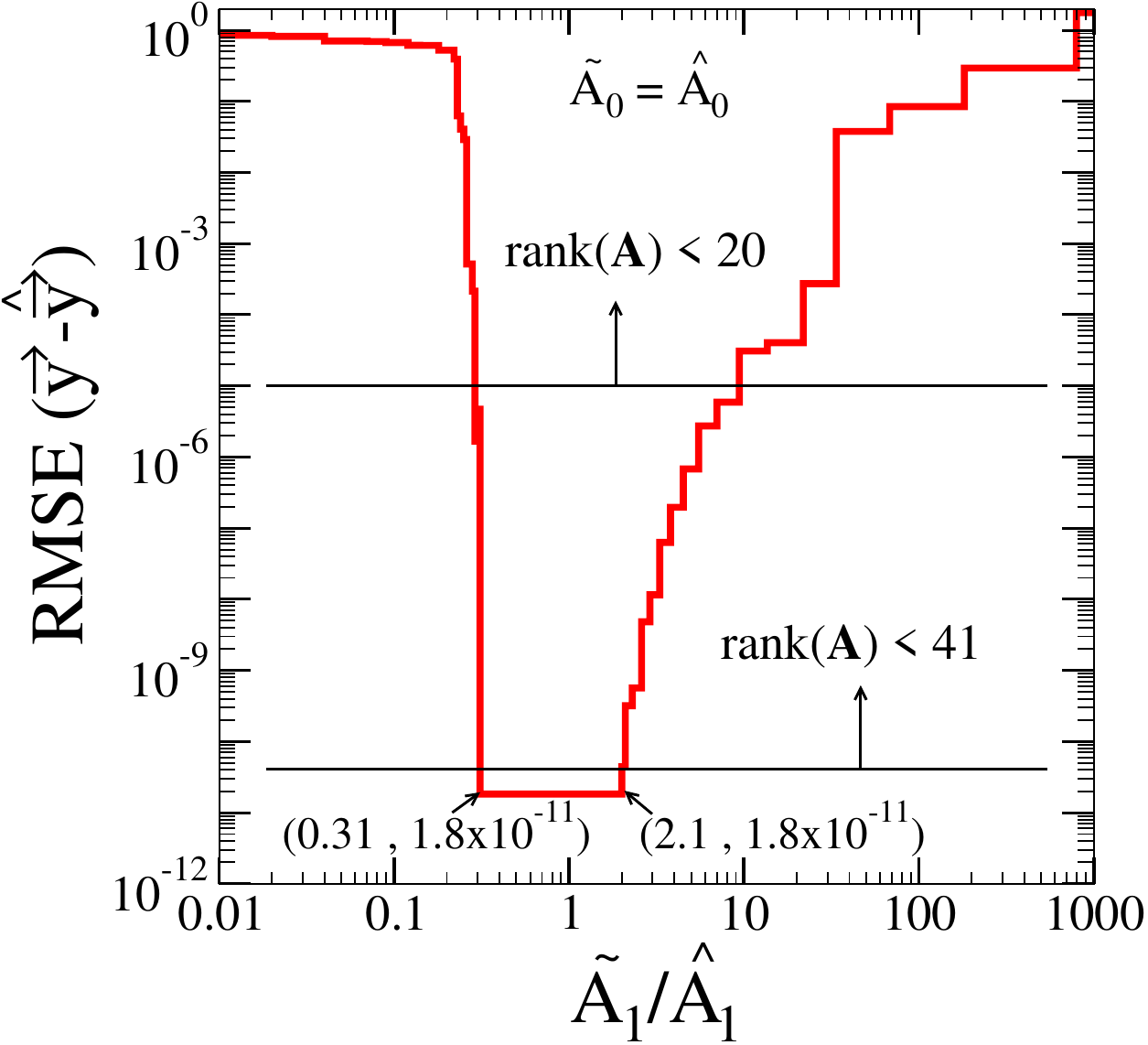}
    \caption{Evaluation of the adequacy of the definition for $\hat{A}_1$. RMSE of the actual values $y(t)$ and the predicted values $\hat{y}(t)$ for different values of $\tilde{A}_1/\hat{A}_1$. Based on input-output data obtained using a single tone in the dynamical variable $x(t)=0.5+1.5\sin(\omega_s t)$.}
    \label{fig:rmseA1}
\end{figure}

Figure~\ref{fig:rmseA1} shows the results of the variation of the $\tilde{A}_1$ parameter and maintaining fixed $\tilde{A}_0$=$\hat{A}_0$.   
For each value of $\tilde{A}_1$, we compute the corresponding parameters $\hat{f}_j$ and $\hat{g}_j$, and use them to evaluate the predicted values for the model ($\hat{\vec{y}}$) from $\vec{x}$. Subsequently, we calculate the Root Mean Square Error (RMSE) of $\vec{y}-\hat{\vec{y}}(t)$, defined as

\begin{equation}
    \text{RMSE}(\vec{y}-\hat{\vec{y}})=\sqrt{\frac{1}{L}\sum_{j=0}^{L-1} (y_j-\hat{y}_j)^2 } \;\; ,
    \label{eq:rmse:clean}
\end{equation}

where $\vec{y}$ corresponds to the values from the DS and $\hat{\vec{y}}$ to the predicted values for the model using $\vec{x}$. 
Figure~\ref{fig:rmseA1} presents the lowest RMSE ($1.8\times 10^{-11}$) for $\tilde{A}_1/\hat{A}_1$ on the interval [$0.31, 2.1$], centered at $\tilde{A}_1$=$1.2\hat{A}_1$. Beyond this interval, the RMSE increases abruptly. 
This behavior can be studied by analyzing the rank of the matrix $\textbf{A}$, which is the matrix that must be inverted to obtain the optimized parameters based on least-squares regression (see Appendix~\ref{app:1}). 
The maximum possible rank for $N$=$20$ is $\text{max}(\text{rank}(\textbf{A}))$=$2N$+1=41, and this value is achieved only in the interval where the RMSE is minimum. 
Conversely, higher RMSE values correspond to $\text{rank}(\textbf{A})\!<$41, indicating that the matrix $\textbf{A}$ loses full rank outside the interval of minimum RMSE.

\begin{figure}[!ht]
    \centering
    \includegraphics[scale=0.4]{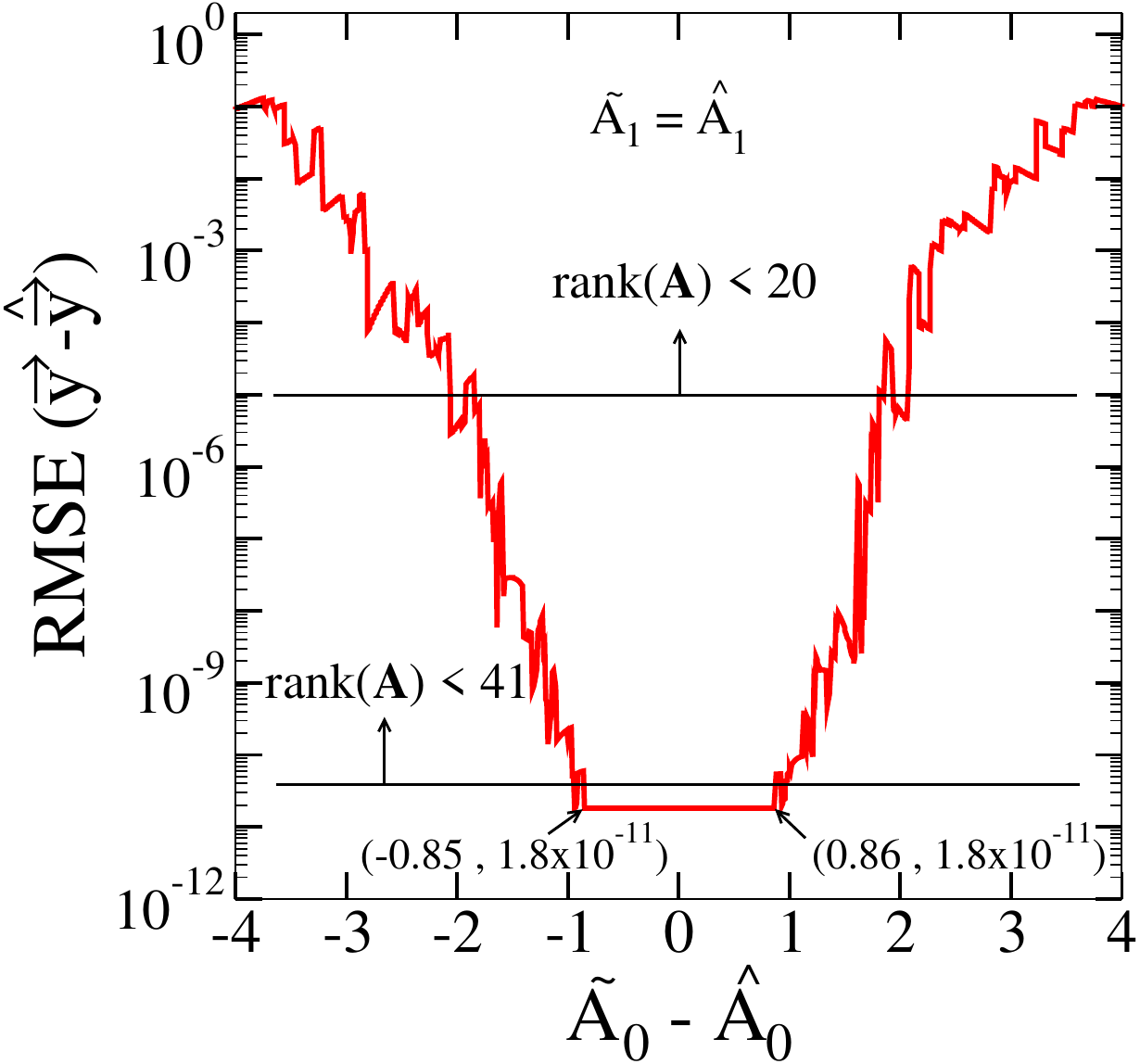}
    \caption{Evaluation of the adequacy of the definition for $\hat{A}_0$. RMSE of the actual values $y(t)$ and the predicted values $\hat{y}(t)$ for different values of $\tilde{A}_0-\hat{A}_0$. Based on input-output data obtained using a single tone in the dynamical variable $x(t)=0.5+1.5\sin(\omega_s t)$.}
    \label{fig:rmseA0}
\end{figure}

Figure~\ref{fig:rmseA0} presents the results of the analysis regarding the variation of the hyperparameter $\tilde{A}_0$ and maintaining fixed $\tilde{A}_1$=$\hat{A}_1$. 
The RMSE($\vec{y}-\hat{\vec{y}}$) is computed for each $\tilde{A}_0$ value following the previously described procedure. 
The minimum RMSE occurs in the interval $\tilde{A}_0\in[\hat{A}_0-0.85, \hat{A}_0+0.86]$, approximately centered at $\tilde{A}_0$=$\hat{A}_0$. These analyses collectively indicate that the definitions of $\hat{A}_0$ and $\hat{A}_1$ from Eq.~\ref{eq:1ordsys:A0A1def} are near the midpoint of the intervals associated with the minimal RMSE values. Moreover, based on the insights gained in this section, these definitions ensure the attainment of the maximum $rank(\textbf{A})$, which was demonstrated for this illustrative example.  Finally, the definitions of $\hat{A}_0$ and $\hat{A}_1$ from Eq.~\ref{eq:1ordsys:A0A1def} are suitable. For the rest of this work, we adopt those definitions for the hyperparameters of model 1.

\subsection{Considerations about the maximum order N}\label{sec:illustrative:maxN}

\begin{figure}[!ht]
    \centering
    \includegraphics[scale=0.4]{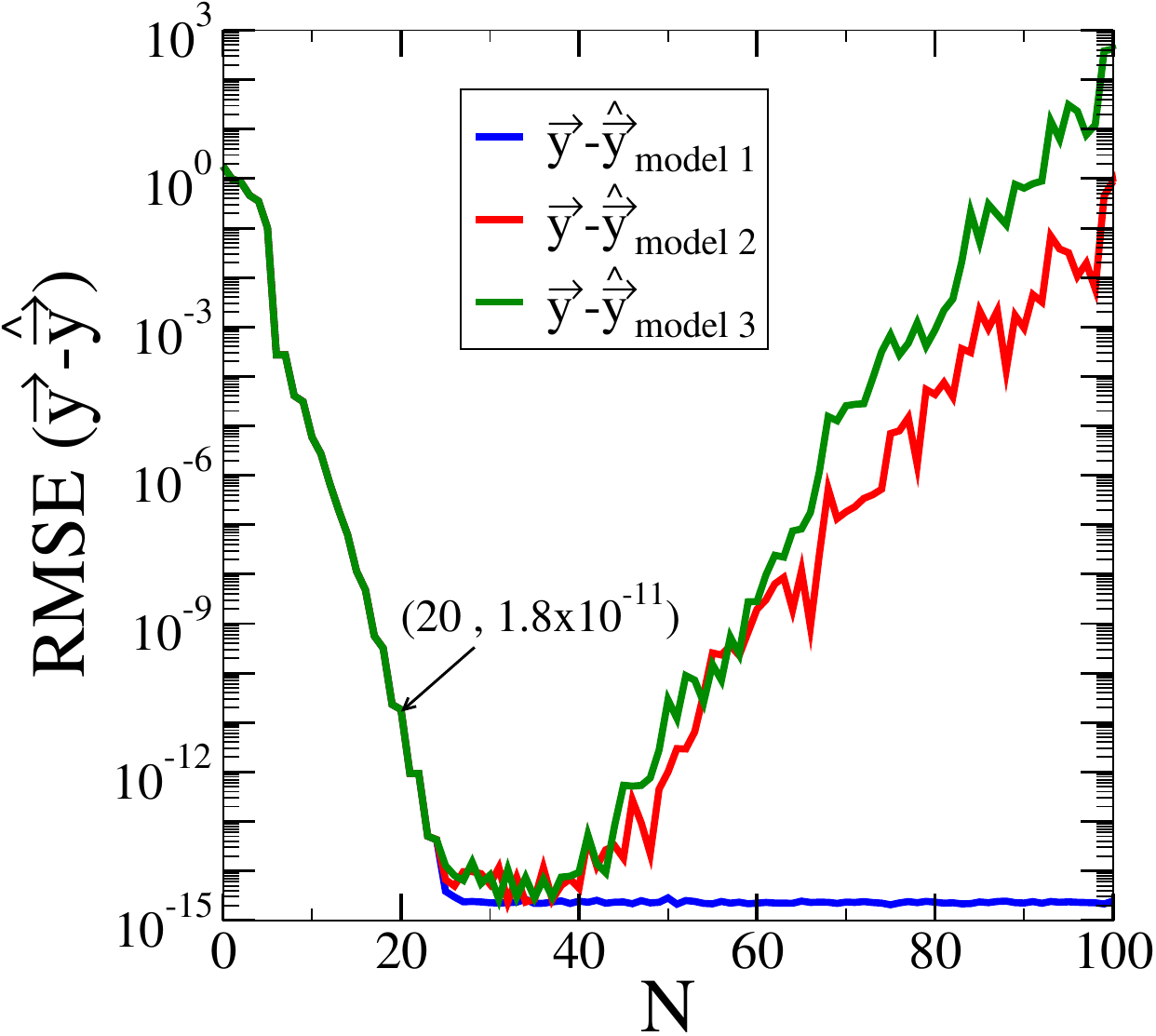}
    \caption{RMSE values as a function of the maximum order $N$ for the three models used in this work. DS obtained by using a single tone in the dynamical variable $x(t)=0.5+1.5\sin(\omega_s t)$.}
    \label{fig:rmseN}
\end{figure}

In this section, we discuss some considerations regarding the influence of the chosen order $N$ in the formalism. As mentioned in Sec.~\ref{sec:formalism} (and further detailed in Sec.~\ref{sec:formalism:procedure}), we perform the system estimation using model 1 and then, based on mathematical equivalences between the models, we use model 1 to compute models 2 and 3. Thus, the predicted values $\hat{y}(t)$ for models 2 and 3 are expected to coincide with those of model 1. For this illustrative example, Fig.~\ref{fig:rmseN} show RMSE values as a function of $N$ for the three models. For $N<$25, the results coincide exactly, indicating greater truncation errors as the order decreases. For $N>$25, the predicted values for model 1 stabilize at $\sim$2.2$\times\!10^{-15}$, while those for models 2 and 3 start increasing at N$\sim$40 due to roundoff errors. 
It is important to note that, although the three models are mathematically equivalent, additional considerations must be taken into account when they are implemented on a computer. The mathematical equalities that relate the three models extensively use the combination formula, which produces increasingly large numbers as $N$ increases. For $N>$40, these values can exceed the maximum limits that standard variable types can represent (whether integer or double). Consequently, the results are rounded off, leading to an increase in RMSE values for models 2 and 3. These roundoff errors can be mitigated by using advanced techniques in variable definitions, such as variable- or arbitrary-precision arithmetic.  However, reducing these roundoff errors would not improve the model in this illustrative example, as the errors are bounded by model 1 and they do not decrease further above $N$=40. Therefore, for practical purposes, it is advisable to use $N$ values from 20 to 50. For this illustrative example, we have chosen $N$=20 because we intend to analyze each parameter individually, and it is simpler to present a smaller number of parameters in tables.

\subsection{Modeling and simulating using different Datasets from a given $x(t)$}\label{sec:illustration:datasets}

In this section, we examine the efficacy of the formalism in identifying different DSs and the comparison between models parameters using models 2 and 3, where these DSs are generated by setting a given $x(t)$ and evaluating the corresponding $y(t)$ using Eq.~\ref{eq:1ordsys}. 
To achieve this, we evaluate the \textbf{r}oot \textbf{m}ean \textbf{s}quare \textbf{e}rror (RMSE) for the predicted values of $\hat{y}(t)$ in comparison to the input-outpu data, i.e. RMSE($y(t)$-$\hat{y}(t)$), and also analyze the errors in the obtained parameter values for the models 2 and 3 for each DS. 
The latter is estimated by calculating RMSE values for the model parameters with respect to these parameters obtained from a reference DS. These analyses aim to enhance our understanding of the behavior of the parameters involved for the different models.

It is important to highlight again that the range for $x(t)$ is possibly different for every DS, resulting in different values for $\hat{A}_0$ and $\hat{A}_1$. 
As a consequence, the elements of $\hat{\vec{f}}$ and $\hat{\vec{g}}$ can also be different for different DSs, making a direct comparison between the model parameters challenging. 
To address this issue, we use model 2. This is obtained by transforming the obtained parameters, given by the vectors $\hat{\vec{f}}$ and $\hat{\vec{g}}$, into parameters corresponding to another power series with hyperparameters $A_0$ and $A_1$, defined by the vectors $\vec{f}$ and $\vec{g}$ (Eq.~\ref{eq:1ordsys:fgpol}). 
In order to compare parameters given by $\vec{f}$ and $\vec{g}$ obtained from different DSs, the hyperparameters $A_0$ and $A_1$ must be the same for all DSs. 

\begin{figure}[!ht]
    \centering
    \includegraphics[scale=0.35]{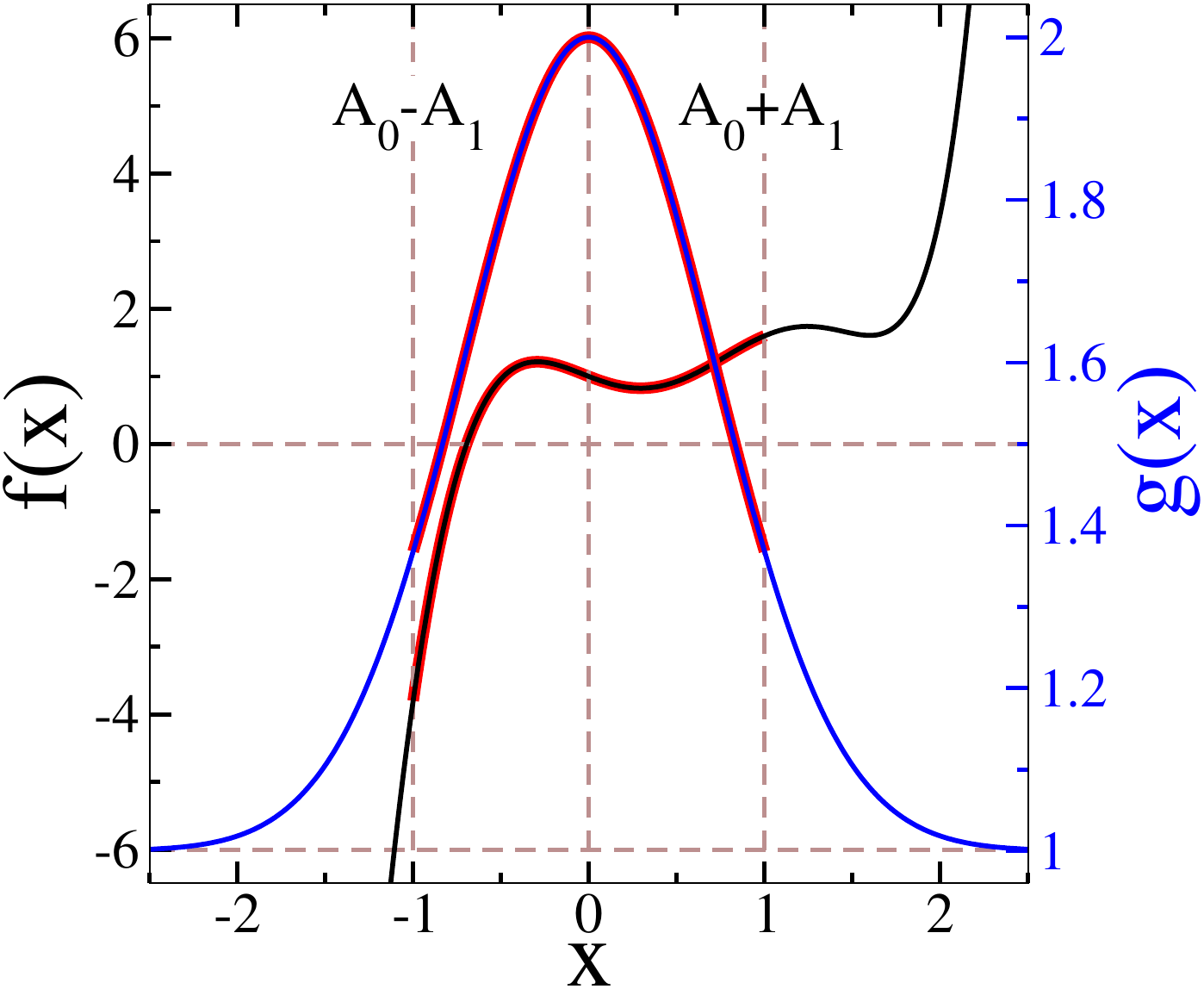}
    \caption{Characteristic curves f(x) and g(x) for the illustrative example, along with the interval [$A_0-A_1,A_0+A_1$] used for models 2 and 3. Black solid line and left axis: $f(x)$, blue solid line and right axis: $g(x)$.}
    \label{fig:fg}
\end{figure}

Moreover, we  compute model 3 by using the obtained parameters from model 2, this also enables us to compare the parameters of the Fourier analysis-based model obtained from different DSs.  
Specifically, we can use the parameters $\vec{f}$ and $\vec{g}$ from model 2 to obtain the vectors $\overrightarrow{\smash{ab}\vphantom{b}}_f$ and $\overrightarrow{\smash{ab}\vphantom{b}}_g$ by using  Eq.~\ref{eq:sirps:vecab2fg}, whose components are the Fourier coefficients $a_j$ and $b_j$, as defined in Eq.~\ref{eq:sirps:defs}. 
With this procedure, we essentially obtain the Fourier coefficients of $y(t)$ that would have if the dynamical variable were $x(t)=A_0+A_1\sin(\omega t)$. These vectors $\overrightarrow{\smash{ab}\vphantom{b}}_f$ and $\overrightarrow{\smash{ab}\vphantom{b}}_g$ also define the characteristic curves completely, because $f(x)$ and $g(x)$ can be expressed as a function of  $\overrightarrow{\smash{ab}\vphantom{b}}_f$, $\overrightarrow{\smash{ab}\vphantom{b}}_g$, and the hyperparameters $A_0$, $A_1$, and $\omega$, according to Eq.~\ref{eq:sirps:vecab2fg}. 
In order to compare parameters of $\overrightarrow{\smash{ab}\vphantom{b}}_f$ and $\overrightarrow{\smash{ab}\vphantom{b}}_g$ for different DSs, the values for hyperparameters $A_0$, $A_1$ and $\omega$ must be the same for all DSs. 
In all calculations henceforth, we employ the hyperparameters $A_0$=0, $A_1$=1, $\omega$=1 and the maximum order considered for the power series is set to $N=20$. Note that the value of $\omega$ does not necessarily have to coincide with $\omega_s$ (for all simulations, we set $T_s$=10 s, $\Delta t$=0.1 s, and therefore, $\omega_s$=2$\pi/100$ s$^{-1}$, but  $\omega$ is always fixed to 1 s$^{-1}$). 
The chosen hyperparameter values for $A_0$ and $A_1$ correspond to an interval $[A_0-A_1,A_0+A_1]$=[-1,1], which can be visualized along with the characteristic curves $f(x)$ and $g(x)$. For this illustrative example, the CCs are shown in Fig.~\ref{fig:fg}, where the interval [-1,1] is represented by the thick red solid line.  
As discussed in Sec.~\ref{sec:formalism:considerations}, we only utilize DSs whose range of $x(t)$ is beyond of the interval $[A_0-A_1,A_0+A_1]$=[-1,1] to avoid incorporating extrapolating errors to the analysis, as we pretend to analyze only the errors arising from the complexity of different DSs.

In certain instances, a model obtained from a given DS may provide accurate predictions when evaluated and compared with the same DS used for modeling, but big discrepancies when it is evaluated with other DSs. Thus, a RMSE error with the same DS used for the fitting is not guarantee the effectiveness of the model. 
To assess this issue, we define a reference DS generated by $x^{\text{ref}}(t)=A_0+A_1\sin(\omega t)$, with $A_0$=0, $A_1$=1 and $\omega$=1, and its corresponding $y^{\text{ref}}(t)$ obtained from Eq.~\ref{eq:1ordsys}. 
The obtained parameters for models 2 and 3 using this reference DS are referred to as $\vec{f}^{\text{ref}}$, $\vec{g}^{\text{ref}}$, $\overrightarrow{\smash{ab}\vphantom{b}}_f^{\text{ref}}$ and $\overrightarrow{\smash{ab}\vphantom{b}}_g^{\text{ref}}$. These reference parameters are useful for comparing the results obtained from the different DSs. 
Furthermore, we can use the models obtained from each DS to evaluate the input $\hat{x}^{\text{ref}}$, obtaining the predicted values $\hat{\vec{y}}^{\text{ref}}$, which can be compared to $\hat{y}^{\text{ref}}$ (this comparison is analogous to validating the model with another DS that was not used during the system modeling).   
Specifically, since $\hat{x}^{\text{ref}}$ is a single tone, this comparison provides a quantitative assessment of how effectively a model identified from a given DS can predict the response of a single tone.

\begin{figure}[!ht]
    \centering
    \includegraphics[scale=0.4]{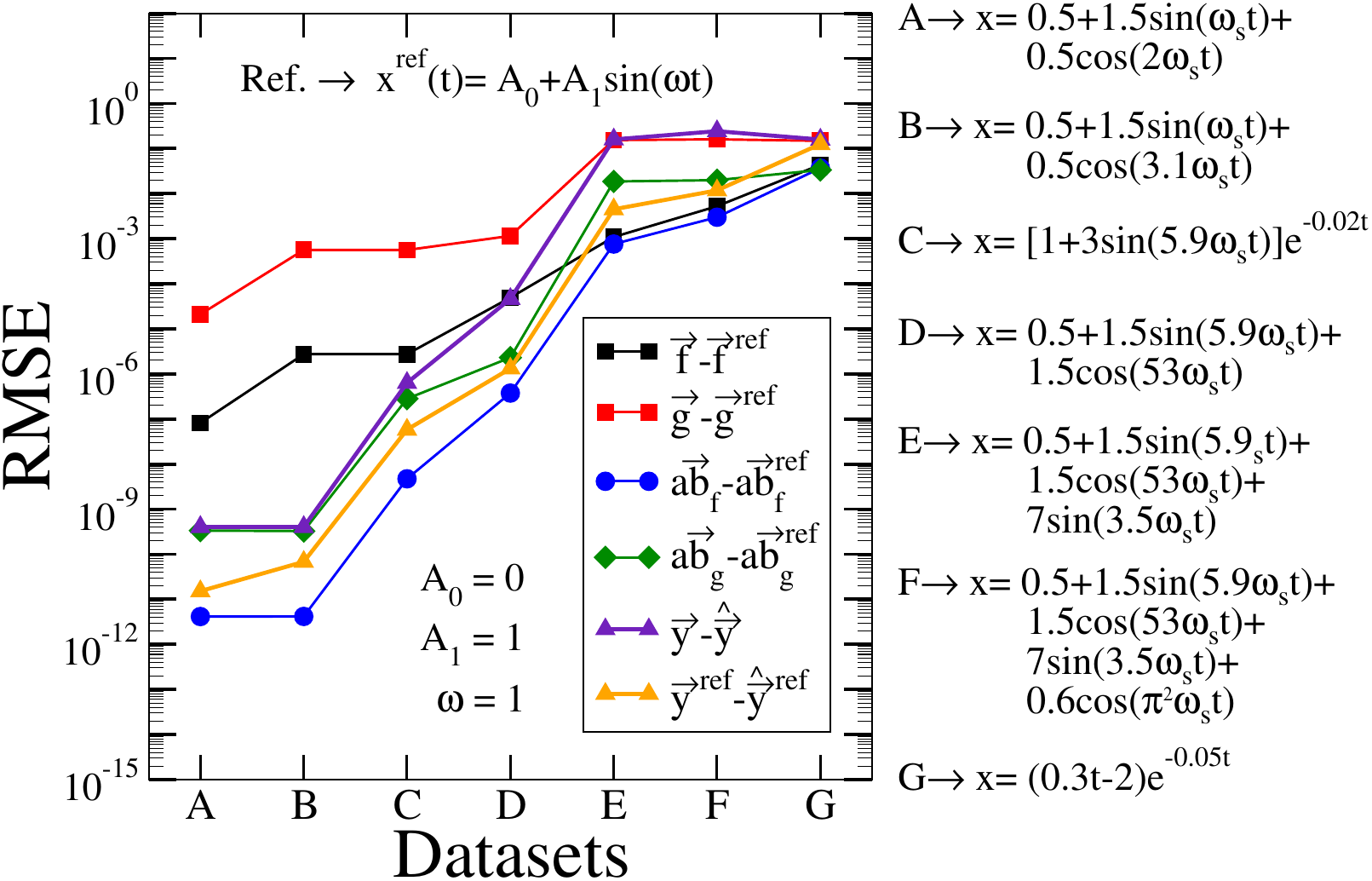}
    \caption{
    RMSE values for different DSs generated by evaluating $y(t)$ with $x(t)$ as indicated on the right. Solid lines are to guide the eye: black (RMSE($\vec{f}-\vec{f}^{~\!\text{ref}}$)), red (RMSE($\vec{g}-\vec{g}^{~\!\text{ref}}$)), blue (RMSE($\vec{ab}_f-\vec{ab}_f^{\text{ref}}$)), green (RMSE($\vec{ab}_g-\vec{ab}_g^{\text{ref}}$)), violet (RMSE($\vec{y}-\hat{\vec{y}}~\!$)), orange (RMSE($\vec{y}^{~\!\text{ref}}-\hat{\vec{y}}^{~\!\text{ref}}$)). Reference parameters are derived from the dynamical variable $x^{\text{ref}}$, and hyperparameters $A_0$ and $A_1$ for all DSs are indicated. Here, $\omega_s=2\pi/T_s$, where $T_s$ is the total simulation time.}
    \label{fig:rmse:x}
\end{figure}

Figure~\ref{fig:rmse:x} illustrates all the mentioned RMSE values using different DSs. 
These DSs, labeled from $A$ to $G$, are generated by defining $x(t)$ as indicated at the right side of the figure and evaluating the corresponding $y(t)$ using Eq.~\ref{eq:1ordsys}. Then, we evaluate the obtained model for the same DS used during the estimation stage, obtaining the predicted values $\hat{\vec{y}}$ for each DS. This allows us to calculate the Root Mean Square Error (RMSE) of $\vec{y}-\hat{\vec{y}}$. 
Similarly, we evaluate the models using the reference DS obtaining $\hat{\vec{y}}^{\text{ref}}$, which allows us to calculate the RMSE values of $\vec{y}^{\text{ref}}-\hat{\vec{y}}^{\text{ref}}$. Both RMSE values exhibit similar trends as a function of the complexity of the DS, displaying an increasing tendency as the DS becomes more complex. This suggests that the least-squares regression system becomes more challenging with the growing complexity of the DS. 
In particular, the RMSE values of $\vec{y}-\hat{\vec{y}}$ for DSs from A to F are approximately one magnitude order greater than those of $\vec{y}^{\text{ref}}-\hat{\vec{y}}^{\text{ref}}$, indicating that the models can predict the single-tone response even better than its own DS used for identification. 
Figure~\ref{fig:rmse:x} also shows the RMSE values for the model parameters using models 2 and 3.  
All these RMSE values are calculated by

\begin{equation}
\begin{aligned}
    \text{RMSE}(\vec{z}-\vec{z}^{\text{ref}}) &= \sqrt{\frac{1}{\text{len}(\vec{z})} \sum_{j=0}^{\text{len}(\vec{z})-1}\left(z_j-z^{\text{ref}}_j\right)^2 } \; \; ,
\end{aligned}
\label{eq:rmsedef}
\end{equation}

where $\vec{z}$ can be any of the vectors $\vec{f}$, $\vec{g}$, $\overrightarrow{\smash{ab}\vphantom{b}}_f$ and $\overrightarrow{\smash{ab}\vphantom{b}}_g$, and $z_j$ is the j-th element of the corresponding vector $\vec{z}$ (where $z_j$ is some parameter from the models). Besides, $\text{len}(\vec{z})$ indicates the number of components of $\vec{z}$, with $\text{len}(\vec{f})=\text{len}(\overrightarrow{\smash{ab}\vphantom{b}}_f)=N+1$ and $\text{len}(\vec{g})=\text{len}(\overrightarrow{\smash{ab}\vphantom{b}}_g)=N$, where $N$ is the maximum order chosen for the polynomial expansion (set here as $N=20$). 

It is interesting to mention that the elements in $\overrightarrow{\smash{ab}\vphantom{b}}_f^{\text{ref}}$ and $\overrightarrow{\smash{ab}\vphantom{b}}_g^{\text{ref}}$ (which correspond to the Fourier coefficients obtained from the reference DS) coincide exactly with the Fourier coefficients calculated of $y(t)$ when $x(t)=A_0+A_1\sin(\omega t)$, as justified by the SORPS formalism. Consequently, the RMSE values of $\overrightarrow{\smash{ab}\vphantom{b}}_f-\overrightarrow{\smash{ab}\vphantom{b}}^{\text{ref}}_f$ and $\overrightarrow{\smash{ab}\vphantom{b}}_g-\overrightarrow{\smash{ab}\vphantom{b}}^{\text{ref}}_g$ can be interpreted as a measure of the distance between the Fourier coefficients of $y(t)$ when we simulate a single tone $x(t)=A_0+A_1\sin(\omega t)$ with the obtained model from each DS and the Fourier coefficients that we effectively obtain when we apply this single tone. This interpretation is possible since the reference DS has been generated for $x(t)$ defined as a single tone with hyperparameters $A_0$, $A_1$ and $\omega$, whose values are the same that those of the hyperparameters used for obtaining model 3. 

As the three models are mathematically equivalent and predict the same results for $y(t)$, we observe that the focus for obtaining a good system estimation is only concentrated to the first model, i.e. to the least-squares regression system. 
We briefly mention that, as the DSs used for system modeling become more complex, the least-squares regression system becomes more problematic. This can be evidenced, for instance, by a greater value for the condition number of the least-squares regression matrix (not shown here), which may be directly related to the observed growing tendency of the RMSE curves. A detailed analysis of this issue deserves more considerations that are out of the scope of the work, and may be addressed elsewhere.

As seen in Fig.~\ref{fig:rmse:x}, for DSs from $A$ to $D$, the RMSE values for $\vec{f}-\vec{f}^{\text{ref}}$ are at least two orders of magnitude greater compared to the values for $\overrightarrow{\smash{ab}\vphantom{b}}_f-\overrightarrow{\smash{ab}\vphantom{b}}_f^{\text{ref}}$. And the corresponding RMSE values being approximately equal for DSs from $E$ to $G$ (but always lower for $\overrightarrow{\smash{ab}\vphantom{b}}_f-\overrightarrow{\smash{ab}\vphantom{b}}^{\text{ref}}_f$).
A similar pattern is observed for RMSE values of $\vec{g}-\vec{g}^{\text{ref}}$ compared to $\overrightarrow{\smash{ab}\vphantom{b}}_g-\overrightarrow{\smash{ab}\vphantom{b}}_g^{\text{ref}}$. They are at least three orders of magnitude greater for DSs from $A$ to $D$ and approximately one magnitude order greater from $E$ to $F$. This suggests that parameters of the power series ($\vec{f}$ and $\vec{g}$) exhibit wider variation compared to the Fourier coefficients ($\overrightarrow{\smash{ab}\vphantom{b}}_f$ and $\overrightarrow{\smash{ab}\vphantom{b}}_g$). 
To support this observation, Table~\ref{table:musigma:x} presents the mean ($\mu$) and standard deviation ($\sigma$) values for each element of the vectors $\vec{f}$, $\vec{g}$, $\overrightarrow{\smash{ab}\vphantom{b}}_f$ and $\overrightarrow{\smash{ab}\vphantom{b}}_g$ (i.e. the model parameters) calculated using all DSs from A to G.  
For each parameter $z$, the values of $\mu(z)$ and $\sigma(z)$ are calculated using the usual formulas, given by

\begin{equation}
\begin{aligned}
 \mu(z)&=\frac{1}{\text{n(DS)}}\sum_{i=1}^{\text{n(DS)}} z_i    \\
 \sigma(z)&=\sqrt{\frac{1}{\text{n(DS)}}\sum_{i=1}^{\text{n(DS)}} \left(z_i-\mu(z)\right)^2}    \; \;,      
\end{aligned}
\end{equation}
where $i$ is an integer that numerates the DSs, $z_i$ indicates the value of the parameter $z$ for the $i$-th DS, and $\text{n(DS)}$ indicates the number of DSs considered in the analysis ($\text{n(DS)}$=7). 
Firstly, it is noted that, since we have chosen $A_0$=0 and $A_1$=1, the values of $\vec{f}$ can be directly related to the theoretical coefficients $f_i^{th}$ of Eq.~\ref{eq:example:charcuves}, resulting in a strong agreement, with the largest error being 10\% for $f_7$.      
At the bottom of Table~\ref{table:musigma:x}, the sum of the standard deviations for all the element of the corresponding vector is represented as $\sum\sigma(z)$. It is observed that the value of $\sum\sigma(z)$ for $\vec{f}$ is greater than the analogous value for $\overrightarrow{\smash{ab}\vphantom{b}}_f$. Furthermore, the value of $\sum\sigma(z)$ for $\vec{g}$ is much greater than the analogous value for $\overrightarrow{\smash{ab}\vphantom{b}}_g$, indicating that the elements of the vector $\vec{g}$ exhibit greater variation. This aligns with the results in Fig.~\ref{fig:rmse:x}, where larger RMSE values are observed for $\vec{f}$ and $\vec{g}$ compared to  $\overrightarrow{\smash{ab}\vphantom{b}}_f$ and $\overrightarrow{\smash{ab}\vphantom{b}}_g$, respectively.
From Table~\ref{table:musigma:x}, it is observed that the $\sigma$ values for parameters from $f_5$ to $f_9$, are of the order of $10^{-2}$, which is one order of magnitude greater than those for parameters from $f_0$ to $f_4$. Similarly, the $\sigma$ values for the parameters $g_2$, $g_4$, $g_6$, $g_7$ and $g_8$ are of the order of $10^{-1}$, while for $g_0$ and $g_1$ are one magnitude order smaller. 
However, for $\overrightarrow{\smash{ab}\vphantom{b}}_f$ and $\overrightarrow{\smash{ab}\vphantom{b}}_g$, the variation of parameters are more concentrated towards the lower index parameters. In particular, the values of $\sigma$ for parameters from $a_0$ to $a_5$ and from $b_1$ to $b_4$ are of the order of $10^{-2}$, while for all the other parameters, their $\sigma$ values are at least one order of magnitude greater.

\begin{table*}[htpb]
\begin{adjustwidth}{-2cm}{-2cm}  
\footnotesize 
\caption{Mean values $\mu(z)$ and standard deviations $\sigma(z)$ for each parameter $z$ of the vectors $\vec{f}$, $\vec{g}$, $\vec{ab}_f$, and $\vec{ab}_g$: evaluating $y(t)$ with varied $x(t)$ functions. DSs: $\text{A}$, $\text{B}$, $\text{C}$, $\text{D}$, $\text{E}$, $\text{F}$, and $\text{G}$.}
\begin{tabular*}{1.25\textwidth}{ 
  @{\extracolsep{\fill}} r 
  S[table-format=1.0e+2, round-mode=places, round-precision=1]
  S[table-format=1.0e+2, round-mode=places, round-precision=0]
  |r
  S[table-format=1.0e+2, round-mode=places, round-precision=1]
  S[table-format=1.0e+2, round-mode=places, round-precision=0]
  |r
  S[table-format=1.0e+2, round-mode=places, round-precision=1]
  S[table-format=1.0e+2, round-mode=places, round-precision=0]
  |r
  S[table-format=1.0e+2, round-mode=places, round-precision=1]
  S[table-format=1.0e+2, round-mode=places, round-precision=0] 
  @{}
}
\toprule
 \multicolumn{3}{c|}{$\vec{f}$} & \multicolumn{3}{c|}{$\vec{g}$} & \multicolumn{3}{c|}{$\overrightarrow{\smash{ab}\vphantom{b}}_f$} & \multicolumn{3}{c}{$\overrightarrow{\smash{ab}\vphantom{b}}_g$} \\
{$z$} & {$\mu(z)$} & {$\sigma(z)$} & {$z$} & {$\mu(z)$} & {$\sigma(z)$} & {$z$} & {$\mu(z)$} & {$\sigma(z)$} & {$z$} & {$\mu$(z)} & {$\sigma(z)$}  \\
\midrule
  $f_0$   & 1.00E+00 & 1.45E-03 & $g_0$   & 1.96E+00 & 5.97E-02 & $a_0$   & 2.59E-01 & 1.96E-02 & $a_1$   & 1.78E+00 & 2.10E-02 \\
  $f_1$   & -1.00E+00 & 6.64E-03 & $g_1$   & -7.62E-03 & 1.21E-02 & $b_1$   & 1.78E+00 & 3.72E-02 & $b_2$   & 8.08E-03 & 2.74E-02 \\
  $f_2$   & 5.02E-01 & 3.33E-03 & $g_2$   & -8.49E-01 & 2.39E-01 & $a_2$   & 1.05E+00 & 3.05E-02 & $a_3$   & 1.67E-01 & 4.91E-02 \\
  $f_3$   & 3.90E+00 & 4.23E-03 & $g_3$   & 7.14E-03 & 8.69E-03 & $b_3$   & -9.04E-01 & 2.21E-02 & $b_4$   & -1.06E-02 & 2.43E-02 \\
  $f_4$   & -3.00E+00 & 4.05E-03 & $g_4$   & 3.76E-01 & 1.82E-01 & $a_4$   & -2.95E-01 & 1.36E-02 & $a_5$   & 9.49E-03 & 1.66E-02 \\
  $f_5$   & -2.05E-01 & 1.10E-02 & $g_5$   & 1.67E-02 & 4.55E-02 & $b_5$   & -1.54E-02 & 7.06E-03 & $b_6$   & 3.83E-03 & 9.48E-03 \\
  $f_6$   & 4.10E-01 & 2.47E-02 & $g_6$   & -1.60E-01 & 1.32E-01 & $a_6$   & -1.37E-02 & 3.02E-03 & $a_7$   & 2.96E-03 & 4.43E-03 \\
  $f_7$   & -1.58E-02 & 3.88E-02 & $g_7$   & 5.68E-02 & 1.38E-01 & $b_7$   & 4.03E-04 & 9.89E-04 & $b_8$   & -5.55E-04 & 1.36E-03 \\
  $f_8$   & 1.70E-02 & 4.17E-02 & $g_8$   & -2.42E-02 & 1.28E-01 & $a_8$   & 8.05E-05 & 1.97E-04 & $a_9$   & -2.18E-05 & 2.30E-04 \\
  $f_9$   & -1.10E-02 & 2.70E-02 & $g_9$   & 2.71E-02 & 6.76E-02 & $b_9$   & 4.59E-06 & 1.12E-05 & $b_{10}$ & -1.64E-05 & 3.97E-05 \\
  $f_{10}$ & 1.89E-03 & 4.65E-03 & $g_{10}$ & -2.92E-03 & 7.01E-03 & $a_{10}$ & 1.11E-05 & 2.71E-05 & $a_{11}$ & -1.48E-05 & 4.47E-05 \\
  $f_{11}$ & 3.60E-03 & 8.82E-03 & $g_{11}$ & -1.48E-02 & 3.68E-02 & $b_{11}$ & -4.50E-06 & 1.10E-05 & $b_{12}$ & 6.24E-06 & 1.54E-05 \\
  $f_{12}$ & -3.39E-03 & 8.31E-03 & $g_{12}$ & 9.91E-03 & 2.23E-02 & $a_{12}$ & -8.38E-07 & 2.05E-06 & $a_{13}$ & 9.37E-07 & 1.96E-06 \\
  $f_{13}$ & 9.46E-04 & 2.32E-03 & $g_{13}$ & -6.12E-04 & 1.32E-03 & $b_{13}$ & -3.91E-08 & 9.60E-08 & $b_{14}$ & 1.90E-07 & 4.76E-07 \\
  $f_{14}$ & 3.80E-04 & 9.32E-04 & $g_{14}$ & -1.99E-03 & 4.64E-03 & $a_{14}$ & -6.38E-08 & 1.56E-07 & $a_{15}$ & 1.04E-07 & 2.44E-07 \\
  $f_{15}$ & -3.60E-04 & 8.81E-04 & $g_{15}$ & 8.33E-04 & 2.01E-03 & $b_{15}$ & 1.41E-08 & 3.44E-08 & $b_{16}$ & -1.30E-08 & 3.11E-08 \\
  $f_{16}$ & 7.08E-05 & 1.73E-04 & $g_{16}$ & 1.92E-05 & 3.34E-05 & $a_{16}$ & 1.21E-10 & 2.97E-10 & $a_{17}$ & 1.91E-09 & 4.47E-09 \\
  $f_{17}$ & 2.40E-05 & 5.89E-05 & $g_{17}$ & -9.72E-05 & 2.35E-04 & $b_{17}$ & 5.53E-10 & 1.35E-09 & $b_{18}$ & -8.08E-10 & 1.96E-09 \\
  $f_{18}$ & -1.45E-05 & 3.55E-05 & $g_{18}$ & 2.49E-05 & 6.13E-05 & $a_{18}$ & 1.16E-10 & 2.84E-10 & $a_{19}$ & -9.51E-11 & 2.34E-10 \\
  $f_{19}$ & 2.56E-06 & 6.28E-06 & $g_{19}$ & -1.92E-06 & 4.76E-06 & $b_{19}$ & -9.78E-12 & 2.39E-11 & $b_{20}$ & 3.67E-12 & 9.07E-12 \\
  $f_{20}$ & -1.40E-07 & 3.42E-07 &           &           &           & $a_{20}$ & -2.66E-13 & 6.51E-13 &           &           &           
     \\ \midrule
  &  & {$\sum\sigma(z)=$} & &  & {$\sum\sigma(z)=$} & &  & {$\sum\sigma(z)=$} & &  & {$\sum\sigma(z)=$}\\
  & & 1.89E-01 & &  & {$1\!\!\times\!\!10^{0}$} & &  & 1.34E-01 & &  & 1.54E-01\\
  \bottomrule
\end{tabular*}
\label{table:musigma:x}
\end{adjustwidth}
\end{table*}

\subsection{Modeling and simulating using different Datasets from a given y(t)}\label{sec:illustration:datasets:y}
\begin{figure}[htpb]
    \centering
    \includegraphics[scale=0.4]{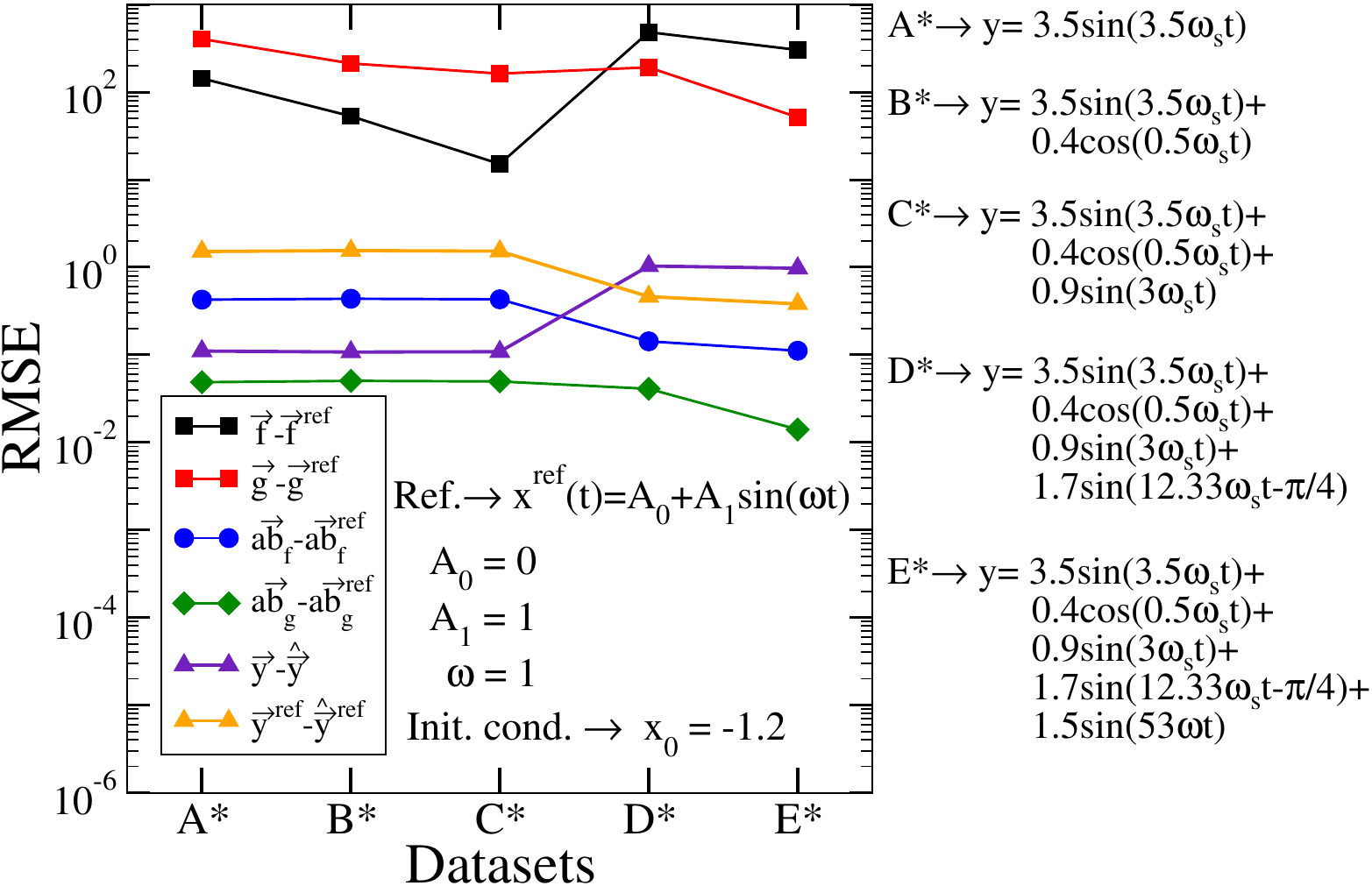}
    \caption{RMSE values for different DSs: numerical integration of $x(t)$ with varied $y(t)$ functions (indicated on the right). Solid lines are to guide the eye: black (RMSE($\vec{f}-\vec{f}^{\text{ref}}$)), red (RMSE($\vec{g}-\vec{g}^{\text{ref}}$)), blue (RMSE($\vec{ab}_f-\vec{ab}_f^{\text{ref}}$)), green (RMSE($\vec{ab}_g-\vec{ab}_g^{\text{ref}}$)), violet (RMSE($\vec{y}-\hat{\vec{y}}$)), orange (RMSE($\vec{y}_{sim}-\hat{\vec{y}}_{sim}$)). Reference parameters are derived from the dynamical variable $x^{\text{ref}}$, initial conditions and hyperparameters $A_0$ and $A_1$ for all DSs are indicated. Here, $\omega_s=2\pi/T_s$, where $T_s$ is the total simulation time.}
    \label{fig:rmse:y}
\end{figure}

For completeness, in this section, we utilize another set of DSs obtained by setting a driven force $y(t)$ and calculating the dynamical variable $x(t)$ through numerical integration. Different driven forces $y(t)$ are defined, and the corresponding dynamical variables $x(t)$ are obtained using the fourth-order Runge-Kutta integration method (RK4) \cite{Runge1895, Kutta1901}.
Figure~\ref{fig:rmse:y} presents the analysis of the estimation process using different DSs labeled from $\text{A}^*$ to $\text{E}^*$. These DSs are generated by setting $y(t)$ as defined on the right side of the figure and by calculating the corresponding $x(t)$ by numerical integration with the initial condition $x_0$=-1.2 for each DS.  
The DSs from $\text{A}^*$ to $\text{E}^*$  begin with a single tone (A$^*$) and progressively incorporate additional terms in an effort to introduce complexity to the estimation process.  
We utilize the same reference DS defined in Sec.~\ref{sec:illustration:datasets}. Furthermore, for all DSs, we have used the hyperparameters $\hat{A}_0$ and $\hat{A}_1$ as defined in Eq.~\ref{eq:1ordsys:A0A1def}, and the hyperparameter values for models 2 and 3 are the same as Sec.~\ref{sec:illustration:datasets}, i.e., $A_0$=0, $A_1$=1 and $\omega$=1.

\begin{figure}
    \centering
    \includegraphics[scale=0.33]{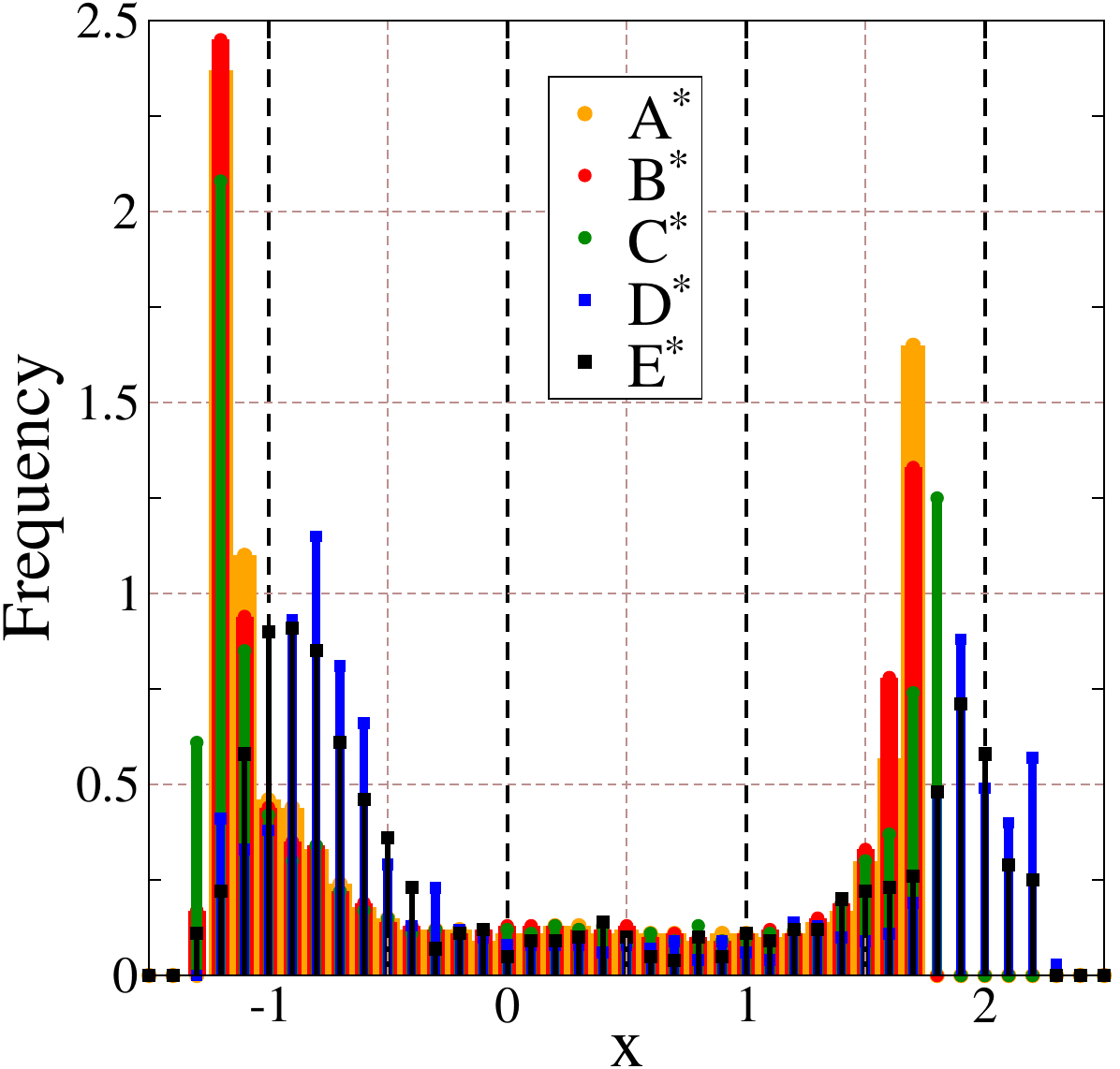}
    \caption{Normalized histograms of $x(t)$ for different DSs. Orange (A$^*$), red (B$^*$), green (C$^*$), blue (D$^*$), black (E$^*$)}
    \label{fig:histogram:y}
\end{figure}

By analyzing the RMSE values in Fig.~\ref{fig:rmse:y}, we observe a growing tendency of $\vec{y}-\hat{\vec{y}}$ as the DS becomes more complex. On the contrary, the  RMSE values of $\vec{y}^{\text{ref}}-\hat{\vec{y}}^{\text{ref}}$ show a decreasing tendency. 
This is a non-intuitive result that deserves some discussion. 
In other words, for the DSs named D$^*$ and E$^*$, the RMSE values obtained when we evaluate the same DS used for modeling are greater than for the other DSs but lower for the reference DS. It indicates that the system is better identified from the complex DSs D$^*$ and E$^*$ compared to the other DSs that are apparently less complex.
This can be further analyzed by representing histogram values of $x(t)$ for each DS, as shown in Fig.~\ref{fig:histogram:y}. It can be seen that for the DSs A$^*$, B$^*$ and C$^*$, the maximum values of distributions are around $-1.2$ and $1.7$, while for D$^*$ and E$^*$ the maximum values are at $-0.85$ and $1.9$. Thus, the relative amount of values that are inside the interval [$A_0-A_1,A_0+A_1$]=$[-1,1]$ is greater for D$^*$ and E$^*$, and this brings, as a consequence, a better prediction for the model with the reference DSs than with its DS used for system modeling (as the models are evaluated only within the range [-1,1]). 
These results suggest that the relative amount of values inside the interval [$A_0-A_1,A_0+A_1$] should be as large as possible to improve the system modeling stage. 

Furthermore, the RMSE values of $\vec{f}-\vec{f}^{\text{ref}}$ and  $\vec{g}-\vec{g}^{\text{ref}}$ are at least two orders or magnitude greater in comparison to the RMSE values of $\overrightarrow{\smash{ab}\vphantom{b}}_f-\overrightarrow{\smash{ab}\vphantom{b}}_f^{\text{ref}}$ and $\overrightarrow{\smash{ab}\vphantom{b}}_g-\overrightarrow{\smash{ab}\vphantom{b}}_g^{\text{ref}}$, respectively. 
This is in agreement with the results found in Sec.~\ref{sec:illustration:datasets}.  
Here, the differences are even greater than in Sec.~\ref{sec:illustration:datasets}, indicating that the errors in the elements of $\vec{f}$ and $\vec{g}$ must be greater than those in $\overrightarrow{\smash{ab}\vphantom{b}}_f$ and $\overrightarrow{\smash{ab}\vphantom{b}}_g$. 
To evaluate this, Table~\ref{table:musigma:y} shows the mean ($\mu$) and standard deviation ($\sigma$) values for each element of the vectors $\vec{f}$, $\vec{g}$, $\overrightarrow{\smash{ab}\vphantom{b}}_f$ and $\overrightarrow{\smash{ab}\vphantom{b}}_g$ across all DSs (analogous to Sec.~\ref{sec:illustration:datasets}). We obtain very large values of $\mu(z)$ for $z$ from $f_6$ to $f_{17}$ and from $g_5$ to $g_{16}$ and $g_{18}$, specifically, values with orders of magnitude greater than $10^1$, while the corresponding values from Sec.~\ref{sec:illustration:datasets} are lower than $10^{-1}$ (Table~\ref{table:musigma:x}). Furthermore, the corresponding values of $\sigma(z)$ are very high (greater than $10^1$). Surprisingly, the values of $\mu(z)$ for the lower index parameters, e.g., for $z$ equal to $f_0$, $f_1$, $g_0$, $g_1$ and $g_2$, are still relatively low and closest to the values of Table~\ref{table:musigma:x}. 
This seems to indicate that, as DS becomes more complex, the errors of the model parameters for the power series-model (corresponding to $\vec{f}$ and $\vec{f}$) move away from the lower index parameters and shift towards parameters with higher indexes.
However, a distinct pattern emerges for parameters of $\overrightarrow{\smash{ab}\vphantom{b}}_f$ and $\overrightarrow{\smash{ab}\vphantom{b}}_g$ (despite being obtained from the same parameters $\vec{f}$ and $\vec{g}$). The higher values of $\mu(z)$ and greater variations $\sigma(z)$ are still concentrated in the lower index parameters for both $\overrightarrow{\smash{ab}\vphantom{b}}_f$ and $\overrightarrow{\smash{ab}\vphantom{b}}_g$ similar to the results of Sec.~\ref{sec:illustration:datasets}. 

In summary, there is a significant discrepancy in the $\sigma$ values for the elements of $\vec{f}$ and $\vec{g}$ compared to those of $\overrightarrow{\smash{ab}\vphantom{b}}_f$ and $\overrightarrow{\smash{ab}\vphantom{b}}_g$, as evident from the direct comparison of their total values ($\sum\sigma(z)$), which show a difference of at least three orders of magnitude.

This result can be qualitatively interpreted based on the mathematical expressions that transform the vectors $\vec{f}$ and $\vec{g}$ to $\overrightarrow{\smash{ab}\vphantom{b}}_f$ and $\overrightarrow{\smash{ab}\vphantom{b}}_g$ (Eq.~\ref{eq:sirps:vecab2fg}). The components of the matrices (\textbf{M})$_{jk}$, as defined in Eqs.~\ref{eq:sirps:ab2f} and ~\ref{eq:sirps:ab2g}, consist of a sum of combinations with higher values as the number of terms increases. Consequently, for higher matrix indexes (i.e. $j,k>>1$), the values for the components of the matrices are increasingly higher. When obtaining the vectors $\overrightarrow{\smash{ab}\vphantom{b}}_f$ and $\overrightarrow{\smash{ab}\vphantom{b}}_g$ from $\vec{f}$ and $\vec{g}$, these matrices must be inverted. Due to their upper-triangle definitions, the values of their inverse (\textbf{M}$^{-1}$)$_{jk}$ decrease as the indexes $j,k$ increase. 
As a consequence, the parameters of $\overrightarrow{\smash{ab}\vphantom{b}}_f$ and $\overrightarrow{\smash{ab}\vphantom{b}}_g$ decrease as the index of the parameters increases.   
This qualitative explanation justifies why the higher values of $\mu$ and $\sigma$ for the elements of $\overrightarrow{\smash{ab}\vphantom{b}}_f$ and $\overrightarrow{\smash{ab}\vphantom{b}}_g$ are more concentrated towards the lower index parameters.

\begin{table*}[!ht]
\begin{adjustwidth}{-2cm}{-2cm}  
\footnotesize
\caption{Mean values $\mu(z)$ and standard deviations $\sigma(z)$ for each parameter $z$ of the vectors $\vec{f}$, $\vec{g}$, $\vec{ab}_f$, and $\vec{ab}_g$: numerical integration of $x(t)$ with varied $y(t)$ functions. DSs: $\text{A}^*$, $\text{B}^*$, $\text{C}^*$, $\text{D}^*$, and $\text{E}^*$.}
\begin{tabular*}{1.25\textwidth}{
  @{\extracolsep{\fill}} r
  S[table-format=1.0e+2, round-mode=places, round-precision=1]
  S[table-format=1.0e+2, round-mode=places, round-precision=0,table-number-alignment=center]
  |r
  S[table-format=1.0e+2, round-mode=places, round-precision=1]
  S[table-format=1.0e+2, round-mode=places, round-precision=0]
  |r
  S[table-format=1.0e+2, round-mode=places, round-precision=1]
  S[table-format=1.0e+2, round-mode=places, round-precision=0]
  |r
  S[table-format=1.0e+2, round-mode=places, round-precision=1]
  S[table-format=1.0e+2, round-mode=places, round-precision=0] @{}
}
\toprule
 \multicolumn{3}{c|}{$\vec{f}$} & \multicolumn{3}{c|}{$\vec{g}$} & \multicolumn{3}{c|}{$\overrightarrow{\smash{ab}\vphantom{b}}_f$} & \multicolumn{3}{c}{$\overrightarrow{\smash{ab}\vphantom{b}}_g$} \\
{$z$} & {$\mu(z)$} & {$\sigma(z)$} & {$z$} & {$\mu(z)$} & {$\sigma(z)$} & {$z$} & {$\mu(z)$} & {$\sigma(z)$} & {$z$} & {$\mu$(z)} & {$\sigma(z)$}  \\
\midrule
  $f_0$   & 8.74E-01 & 7.69E-02 & $g_0$   & 1.93E+00 & 4.41E-02 & $a_0$   & 8.15E-01 & 4.02E-01 & $a_1$   & 1.75E+00 & 5.39E-02 \\
  $f_1$   & -5.34E-01 & 8.35E-01 & $g_1$   & -4.20E-01 & 4.88E-01 & $b_1$   & 1.02E+00 & 5.52E-01 & $b_2$   & -1.17E-01 & 4.33E-02 \\
  $f_2$   & 3.56E+00 & 4.89E+00 & $g_2$   & -2.52E-01 & 1.56E+00 & $a_2$   & 3.02E-01 & 4.08E-01 & $a_3$   & 1.81E-01 & 5.23E-02 \\
  $f_3$   & 7.78E+00 & 1.01E+01 & $g_3$   & -1.48E+00 & 6.51E+00 & $b_3$   & -3.72E-01 & 2.15E-01 & $b_4$   & -7.14E-02 & 7.08E-02 \\
  $f_4$   & -2.71E+01 & 5.79E+01 & $g_4$   & -9.24E+00 & 2.40E+01 & $a_4$   & -1.71E-01 & 1.29E-01 & $a_5$   & -9.28E-03 & 1.84E-02 \\
  $f_5$   & -3.34E+01 & 5.87E+01 & $g_5$   & 2.43E+01 & 5.07E+01 & $b_5$   & -5.13E-02 & 1.23E-01 & $b_6$   & 6.87E-03 & 7.59E-03 \\
  $f_6$   & 1.11E+02 & 2.33E+02 & $g_6$   & 5.40E+01 & 1.32E+02 & $a_6$   & -4.34E-02 & 4.28E-02 & $a_7$   & -2.18E-03 & 1.99E-02 \\
  $f_7$   & 7.38E+01 & 2.04E+02 & $g_7$   & -1.35E+02 & 2.11E+02 & $b_7$   & 2.82E-02 & 5.53E-02 & $b_8$   & 1.27E-02 & 3.16E-02 \\
  $f_8$   & -2.58E+02 & 4.56E+02 & $g_8$   & -1.19E+02 & 3.28E+02 & $a_8$   & 3.91E-05 & 3.96E-03 & $a_9$   & 1.70E-02 & 2.71E-02 \\
  $f_9$   & -5.66E+01 & 4.19E+02 & $g_9$   & 3.62E+02 & 5.16E+02 & $b_9$   & -1.53E-02 & 1.94E-02 & $b_{10}$ & -9.07E-03 & 2.15E-02 \\
  $f_{10}$ & 3.37E+02 & 4.58E+02 & $g_{10}$ & 7.59E+01 & 4.00E+02 & $a_{10}$ & -1.19E-02 & 4.47E-02 & $a_{11}$ & -2.22E-03 & 8.12E-03 \\
  $f_{11}$ & -3.02E+01 & 5.02E+02 & $g_{11}$ & -5.07E+02 & 7.18E+02 & $b_{11}$ & -1.86E-02 & 2.12E-02 & $b_{12}$ & -3.47E-03 & 1.73E-02 \\
  $f_{12}$ & -2.51E+02 & 2.27E+02 & $g_{12}$ & 8.69E+01 & 3.08E+02 & $a_{12}$ & -1.33E-02 & 3.40E-02 & $a_{13}$ & -8.37E-03 & 1.63E-02 \\
  $f_{13}$ & 8.03E+01 & 3.30E+02 & $g_{13}$ & 3.72E+02 & 5.33E+02 & $b_{13}$ & -3.29E-03 & 9.21E-03 & $b_{14}$ & 4.06E-03 & 6.38E-03 \\
  $f_{14}$ & 1.01E+02 & 9.71E+01 & $g_{14}$ & -1.62E+02 & 2.53E+02 & $a_{14}$ & -4.08E-03 & 6.69E-03 & $a_{15}$ & -1.77E-03 & 4.02E-03 \\
  $f_{15}$ & -5.39E+01 & 1.02E+02 & $g_{15}$ & -1.26E+02 & 1.85E+02 & $b_{15}$ & 3.13E-04 & 3.21E-03 & $b_{16}$ & 1.88E-03 & 2.70E-03 \\
  $f_{16}$ & -1.74E+01 & 5.13E+01 & $g_{16}$ & 8.98E+01 & 1.29E+02 & $a_{16}$ & -4.60E-04 & 3.70E-04 & $a_{17}$ & 2.47E-04 & 4.47E-04 \\
  $f_{17}$ & 1.58E+01 & 8.76E+00 & $g_{17}$ & 7.83E+00 & 1.56E+01 & $b_{17}$ & 1.16E-04 & 2.58E-04 & $b_{18}$ & 1.79E-04 & 2.61E-04 \\
  $f_{18}$ & -3.81E-01 & 9.16E+00 & $g_{18}$ & -1.73E+01 & 2.46E+01 & $a_{18}$ & -1.01E-05 & 7.89E-05 & $a_{19}$ & 6.61E-05 & 9.39E-05 \\
  $f_{19}$ & -1.74E+00 & 2.70E+00 & $g_{19}$ & 3.46E+00 & 5.00E+00 & $b_{19}$ & 6.62E-06 & 1.03E-05 & $b_{20}$ & -6.61E-06 & 9.53E-06 \\
  $f_{20}$ & 3.40E-01 & 3.08E-01 &           &           &           & $a_{20}$ & 6.48E-07 & 5.87E-07 &           &           &           
           
         \\ \midrule
  &  & {$\sum\sigma(z)=$} & &  & {$\sum\sigma(z)=$} & &  & {$\sum\sigma(z)=$} & &  & {$\sum\sigma(z)=$}\\
  & & 3.23E+03 & &  & 3.84E+03 & &  & 2.07E+00 & &  & 4.02E-01\\
\bottomrule
\end{tabular*}
\label{table:musigma:y}
\end{adjustwidth}
\end{table*}

\subsection{System identification in the presence of noise}\label{sec:illustration:noise}
In this final analysis, aimed at complementing the results obtained so far, we briefly explore the system identification in the presence of noise. Specifically, we only utilize the DS named as $\text{A}^*$ but introduce an additive noise term, denoted as $N(t)$. Thus, the DS is generated by defining $\vec{y}=3\sin(3.5\omega_s\, \vec{t})+\vec{N}$, where $\omega_s=2\pi/T_s$ and $T_s$ represents the total simulation time set to $100$~s, with a simulation step of $\Delta t=0.1$~s. Both $\vec{x}$ and $\vec{y}$ vectors have a length of $L=1001$, where $t$ ranges from $0$ to $T_s$. The values of $x(t)$ are computed through the numerical integration of Eq.~\ref{eq:1ordsys} using the RK4 integration method, analogous to Sec.~\ref{sec:illustration:datasets:y}.
Besides, $\vec{N}$ represents an Additive White Gaussian Noise (AWGN), defined by

\begin{equation}
    \vec{N}=\vec{\mathcal{N}}(0,1) \frac{\text{RMSE}( \vec{y}^{\text{without-noise}} )}{\sqrt{\text{SNR}_{\text{desired}}}} \; ,
    \label{eq:Noise}
\end{equation}

where $\vec{\mathcal{N}}(0,1)$ is a random vector composed by $L$ numbers generated from a normal distribution with mean $\mu$=0 and standard deviation $\sigma$=1. The vector  $\vec{y}^{\text{without-noise}}$ corresponds to the components of $y(t)$ without noise, i.e. $\vec{y}^{\text{without-noise}}=\sigma( 3\sin(3.5\omega_s\, \vec{t}) )$, and the RMSE values are calculated using Eq.~\ref{eq:rmse:clean}. Additionally, $\text{SNR}_{\text{desired}}$ denotes the desired value of Signal-to-Noise Ratio (SNR). 
This definition ensures that the value of $\text{SNR}_{\text{desired}}$ coincides with the actual definition of SNR. 
To justify this claim, we can refer to the definition of SNR

\begin{equation}
 \text{SNR}\coloneqq\frac{\text{P}(\vec{y}^{\text{without-noise}})}{\text{P}(\vec{N})}    \; ,
 \label{eq:SNRdef}
\end{equation}
where $\text{P}(\vec x)$ represents the power of the vector $x$, defined as $\text{P}(\vec x)\coloneqq\left(\text{RMSE}(\vec x)\right)^2$. If we substitute this definition into Eq.~\ref{eq:SNRdef}, we obtain:

\begin{equation}
 \text{SNR}\coloneqq\frac{\text{P}(\vec{y}^{\text{without-noise}})}{\text{P}(\vec{N})}=\frac{\left(\text{RMSE}(\vec{y}^{\text{without-noise}})\right)^2}{\left(\text{RMSE}(\vec{N})\right)^2}    \; ,
 \label{eq:SNRdef:rmse}
\end{equation}

and by substituting Eq.~\ref{eq:Noise} into Eq.~\ref{eq:SNRdef:rmse}, we finally obtain

\begin{equation}
\begin{aligned}
 \text{SNR} &=\frac{\left(\text{RMSE}(\vec{y}^{\text{without-noise}})\right)^2}{\left(\text{RMSE}\left(\vec{\mathcal{N}}(0,1) \frac{\text{RMSE}( \vec{y}^{\text{without-noise}} )}{\sqrt{\text{SNR}_{\text{desired}}}}\right)\right)^2} \\ &=\frac{\left(\text{RMSE}(\vec{y}^{\text{without-noise}})\right)^2}{\text{RMSE}\left(\vec{\mathcal{N}}(0,1)\right)^2 \frac{\left(\text{RMSE}( \vec{y}^{\text{without-noise}} )\right)^2}{\text{SNR}_{\text{desired}}}} \\ &=\text{SNR}_{\text{desired}}
    \; ,
 \label{eq:SNRdef:rmse:dem}
\end{aligned}
\end{equation}
where in the second equality, scalar numbers are factored out, and in the third equality, we use the fact that the RMSE of a random vector generated from a normal distribution with a zero mean and standard deviation $\sigma$ is equal to $\sigma$, which, in this case, is $\sigma$=1. Hence, by setting a desired value for $\text{SNR}_{\text{desired}}$ in Eq.~\ref{eq:Noise}, we obtain a noise that is consistent with the definition of SNR.  Typically, the SNR is expressed in decibels ($\text{SNR}_{\text{dB}}$), where $\text{SNR}_{\text{dB}}\coloneqq 10 \log_{10}(\text{SNR})$. 

\begin{figure}[!ht]
    \centering
    \includegraphics[scale=0.35]{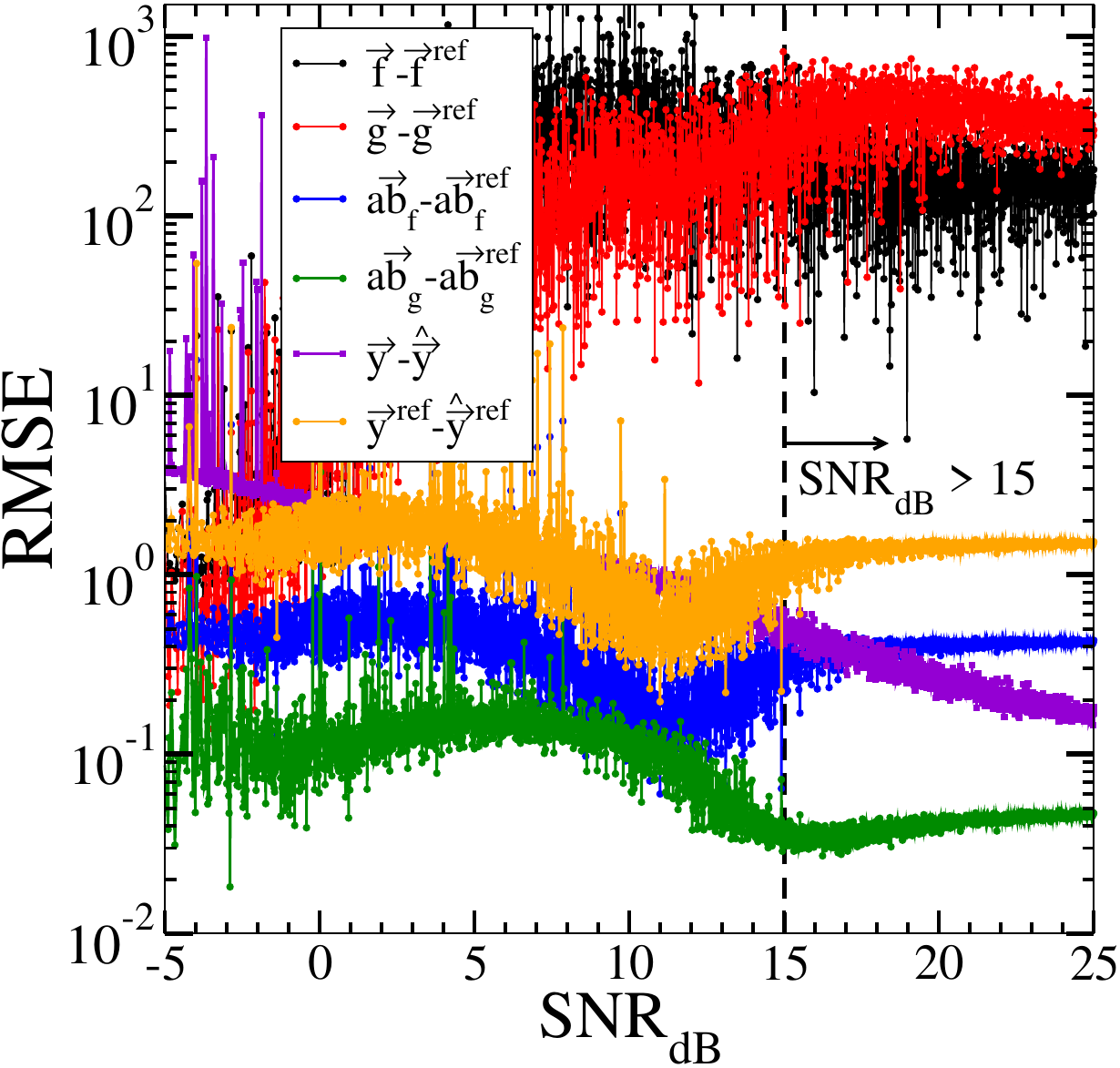}
    \caption{RMSE for system identification in noise presence: DS generated by numerical integration of $x(t)$ for $y(t)=3+\sin(3.5\omega t)+\mathcal{N}(t)$. Solid lines are to guide the eye: black (RMSE($\vec{f}-\vec{f}^{\text{ref}}$)), red (RMSE($\vec{g}-\vec{g}^{\text{ref}}$)), blue (RMSE($\vec{ab}_f-\vec{ab}_f^{\text{ref}}$)), green (RMSE($\vec{ab}_g-\vec{ab}_g^{\text{ref}}$)), violet (RMSE($\vec{y}-\hat{\vec{y}}$)), orange (RMSE($\vec{y}_{sim}-\hat{\vec{y}}_{sim}$)). Reference parameters are derived from the dynamical variable $x^{\text{ref}}$, initial condition $x_0=-1.2$, and hyperparameters $A_0=0$, $A_1=1$, and $\omega=1$.}
    \label{fig:noise}
\end{figure}

Figure~\ref{fig:noise} shows the results of the RMSE for the four vectors $\vec{f}-\vec{f}^{\text{ref}}$, $\vec{g}-\vec{g}^{\text{ref}}$, $\overrightarrow{\smash{ab}\vphantom{b}}_f-\overrightarrow{\smash{ab}\vphantom{b}}_f^{\text{ref}}$ and $\overrightarrow{\smash{ab}\vphantom{b}}_g-\overrightarrow{\smash{ab}\vphantom{b}}_g^{\text{ref}}$ for different values of $\text{SNR}_{\text{dB}}$ ranging from  $-5$ to $25$ ($\text{SNR}$ from 0.316 to 316). 
Here, the reference parameters are obtained from the reference DS, as explained in Sec.~\ref{sec:illustration:datasets}, i.e. they are generated for $x(t)=A_0+A_1\sin(\omega t)$ by evaluating Eq.~\ref{eq:1ordsys}, with $A_0$=0, $A_1$=1 and $\omega$=1.

On the one hand, the RMSE values of $\overrightarrow{\smash{ab}\vphantom{b}}_f-\overrightarrow{\smash{ab}\vphantom{b}}_f^{\text{ref}}$ and $\overrightarrow{\smash{ab}\vphantom{b}}_g-\overrightarrow{\smash{ab}\vphantom{b}}_g^{\text{ref}}$ exhibit a non stable behavior at low $\text{SNR}_{\text{dB}}$, stabilizing for $\text{SNR}_{\text{dB}}>15$ ($\text{SNR}>31.62$), converging to noise-free values, in concordance to the results for the DS named A$^*$ defined in Sec.~\ref{sec:illustration:datasets:y} (Fig.~\ref{fig:rmse:y}). 
On the other hand, the RMSE values of $\vec{f}^{\text{ref}}$ and $\vec{g}-\vec{g}^{\text{ref}}$ also approach to the noise-free values but exhibit more variation, remaining unstable even above of $\text{SNR}_{\text{dB}}\approx 25$ ($\text{SNR}\approx$316).

In summary, the elements of the vectors $\overrightarrow{\smash{ab}\vphantom{b}}_f$ and $\overrightarrow{\smash{ab}\vphantom{b}}_g$ stabilize when the signal power is 31.62 times greater than the noise power. However, the elements of $\vec{f}$ and $\vec{g}$ show more variation even beyond $\text{SNR}\approx 316$.
Similar trends are observed for RMSE values of $\vec{y}-\hat{\vec{y}}$ and $\vec{y}^{\text{ref}}-\hat{\vec{y}}^{\text{ref}}$, where $\vec{y}^{\text{ref}}$ stabilizes for $\text{SNR}_{\text{dB}}>15$, and $\vec{y}$ converges more slowly to the noise-free value of $10^{-1}$.
Consistent with the findings in Secs.~\ref{sec:illustration:datasets} and \ref{sec:illustration:datasets:y}, these results suggest that elements of vectors $\overrightarrow{\smash{ab}\vphantom{b}}_f$ and $\overrightarrow{\smash{ab}\vphantom{b}}_g$ (model 3) exhibit greater robustness and less variation compared to those of $\vec{f}$ and $\vec{g}$ (model 2). Therefore, parameters of model 3 are more useful for being used in comparing the models obtained from different DSs.

\section{Discussion}\label{sec:discussion}

The \textbf{s}inusoidal  \textbf{o}utput \textbf{r}esponse in \textbf{p}ower \textbf{s}eries (SORPS) formalism (as initially introduced in ref.~\cite{Gonzalez2023}), establishes a mathematical connection between parameters of a Fourier analysis-based model and those of a power series-based model (refer to Appendix~\ref{app:2} for details). However, the system identification procedure discussed in that work (based on FFT) presents several limitations: (i) it requires a sinusoidal single tone for the dynamical variable and equally spaced time steps for performing system identification; (ii) it uses hyperparameters $A_0$, $A_1$ and $\omega$ that depend on the input, making it difficult to compare model parameters obtained from different DSs; (iii) There is no clear path for extending the formalism to systems with a higher order than one; and (iv) there is an extrapolation issue that arises when attempting to simulate the system with input data that extends beyond the range used for system identification.

In this work, motivated by the mathematical connection between Fourier analysis- and power series-based models provided by the SORPS formalism, we have presented a different approach for the system identification procedure.  Instead of relying on the Fourier analysis-based model using FFT, we use the power series-based model employing least-squares regression for system identification (referred to as model 1). As a consequence, the system identification procedure no longer requires the use of single tones and equally spaced time steps, thus addressing limitation (i). This is a significant advancement, as it enables us to obtain the system modeling using multi-tone input signals. Furthermore, we propose the utilization of two additional models (models 2 and 3) to facilitate the comparison of parameters obtained from different DSs, thereby addressing limitation (ii). While we have briefly discussed extrapolation issues to tackle limitation (iv), further considerations, along with a more exact definition of ``the complexity of a DS'' is warranted to fully address this issue. 

This work represents a groundbreaking contribution by elevating the concept of characteristic curves (CCs) to the forefront of the formalism. This concept can be straightforwardly applied to higher-order systems, providing a clearer direction for addressing limitation (iii) in future investigations.    
As the CCs completely describe the system, we can parametrize them using different basis functions and select the most suitable parametrization for each particular system. We have demonstrated in this study, based on a first-order system, that the power series-based model (model 1) performs better system identification than the Fourier analysis-based model (as the power series-based model solves many limitations that are present in the Fourier analysis-based model). Hence, we perform system estimation with model 1, and then use model parameters from model 1 to obtain model parameters from models 2 and 3. While these three models are mathematically equivalent (i.e., they predict the same output values from a given input), their individual values differ. Our analysis reveals that when comparing individual parameter values obtained from different DSs, employing the Fourier analysis-based model (model 3) is advantageous. This was evidenced by an illustrative example, which showed smaller RMSE values for model 3 compared to model 2 across all analyses performed.

Finally, further research directions are listed in the following:

(i) Since system modeling relies on least-squares regression (using model 1), future efforts to improve the system modeling process should be focused solely on the least-squares regression problem (as detailed in Appendix~\ref{app:1}). 
In this context, conducting a thorough analysis of the rank and condition number for the least-squares regression matrix under various input-output DSs is an interesting avenue for future research. Particularly, exploring which types of functions and their relative distribution of values should have the input data to improve the system identification process. 

(ii) Additionally, further studies can be conducted on extrapolation issues, which arise when simulating the models beyond the range of data values used during the system identification stage.

(iii) While the SORPS formalism conceptually differs from methods such as OFRF, NOFRF, and HOSIDF, as mentioned in Sec.~\ref{sec:introduction}, it is worthwhile to investigate potential underlying connections. 
Furthermore, the SORPS formalism provides mathematical insights useful for understanding the applications and limitations of Fourier analysis in nonlinear systems, which may complement other techniques based on Fourier analysis. Considerations about the limitations of extending the SORPS formalism to higher-order systems are addressed in Appendix~\ref{app:4}. 
Additionally, since many system estimation techniques, including this work, utilize least-squares regression, further connections with other similar methods could be explored.

(iv) An important future research direction is the use of the concept of CCs for system modeling of higher-order systems.  
It is important to note that most present modeling techniques focus on defining and obtaining models with as few parameters as possible, following the NARMAX-philosophy (which is based on the parsimonious principle\cite{BILLINGS1989Tsang,Pearson1999,Billings_2013}). However, the concept of CCs discussed here call into question this philosophy. Here, the emphasis is placed on the CCs as a whole rather than on the individual values of their parameters. Thus, the use of CCs allows us to define a unique system equation that can describe a wide variety of systems. 
In the particular case of this work, we defined the first-order system of Eq.~\ref{eq:1ordsys}, which represents all the systems containing two variables related by a first-order system. This has the advantage that algebraic operations can be performed beforehand without further considerations for a specific system. 
The logical next step involves defining nonlinear differential equations (NDEs) for higher-order systems based on CCs, starting with a second-order NDE and progressing to higher orders. Following this approach, we propose generating a list of NDEs consisting of first- and higher-order systems. These NDEs can then be systematically tested using a given DS, aiming to identify the lowest order NDE that accurately represents this DS. 
Subsequently, the extension of the parsimonious principle can be concisely stated as obtaining the model with the lowest order that accurately represents the input-output data.  
This concept has the potential to find applications in various areas, including vibration analysis and structural dynamics\cite{nayfeh1979nonlinear,Nayfeh_2000,Nayfeh_2004,Kerschen2006,NOEL20172}, viscoelastic materials\cite{Zhou2020,Hyun2011,banks2011,Younesian2019,SHU2022,Khaniki2022}, design and modeling of nonlinear electric circuits\cite{Narendra1990,slotine1991,Ljung1999,Pintelon2012,Vidyasagar_2002,Khalil2002,Ljung2019}, voltammetry techniques in electrochemistry\cite{Fasmin_2017,Orazem_2017,Wolff_2019,VIDAKOVICKOCH2021100851,Wang_2021}, and structural
health monitoring and fault diagnosis\cite{Jing2015,Zhao2015,Zhao2024}.
Exploring these applications emerges as a natural direction for future research.

\section{Conclusions}\label{sec:conclusion}

Based on the previously reported connection between power series and Fourier series, we have modified the system identification process, which was previously based on \textbf{f}ast \textbf{F}ourier \textbf{t}ransform (FFT), by a least-squares regression method. 
This allows us to obtain the system model from arbitrary input-output data, providing a more practical and widely applicable approach for real-world systems.  
A thorough analysis of the selection of hyperparameters $\hat{A}_0$ and $\hat{A}_1$ for the power series-based model used during the identification process, reveals suitable definitions for these parameters. 
Furthermore, the transformation from the power series-based model used for modeling to another power series-based model but with hyperparameters $A_0$ and $A_1$ fixed for all the DSs, facilitates the comparison of parameters across different DSs.  
Also, the mathematical connection between power series and Fourier series, allows us to use the power series-based model parameters to obtain the Fourier analysis-based model parameters that we would have if we apply a single sinusoidal tone in the dynamical variable. The three models discussed in this work represent the same characteristic curves, as they are mathematically equivalent, but their parameter values differ. 
A comparative analysis of these parameters for the power series- and Fourier analysis-based models was conducted using an illustrative example, where they were analyzed across different DSs and in the presence of noise, revealing noteworthy insights.
Parameters of Fourier analysis-based model displays notably less variation compared to parameters of the power series-based model. This finding emphasizes the use of the Fourier analysis-based model as a more reliable for comparing model parameters obtained from different DSs.

Finally, while the SORPS methodology is rigorously applicable solely to first-order NDEs (or to systems that can be expressed as a set of first-order NDEs), the approach outlined in this work represents a fundamental contribution to the develop of system identification techniques based on the concept of CCs. 
This concept is presented as an alternative or complement to the commonly used approach that emphasizes the estimation of the individual parameter values of the model. Instead, the CCs-based methods emphasize the computation of the CCs as a whole. 
Thus, the parsimonious principle defined by the NARMAX-philosophy is changed from the concept of a model with as few parameters as possible to the concept of finding the lowest model order that correctly describe the input-output data.   
CCs-based models present the advantages that the system identification is uniquely defined, and that it can be applied for every system without any additional algebraic operations.   
The concept of CCs is broad, and numerous system identification techniques, including NOFRF, OFRF, HOSIDFs, sparse regression and HBM-based methods may be benefited from it.


\bmhead{Acknowledgments}
This work has been supported by the ANPCyT Project PICT-2021-I-A-01135, CONICET Project PIP 1679, and the UNR Project PID 80020190100011UR (Argentina).
The author thanks H.F.Busnengo, L.Manuel, J.C.G{\'o}mez, B.J.A.G{\'o}mez and S.S.Quispe-Estrella for valuable discussions.
\bmhead{Competing interests}
The author declares no competing interests.

\bmhead{Data availability}
Data sets generated during the current study are available from the corresponding author on reasonable request. 

\noindent

\begin{appendices}

\section{Least-squares regression system} \label{app:1}

This appendix is dedicated to explain how parameters $\hat{f}_j$ and $\hat{g}_j$ can be calculated based on least-squares regression method. From Eqs.~\ref{eq:1ordsys:fgtilde} and \ref{eq:1ordsys} we have

\begin{equation}
\begin{aligned}
 y(t) = &\sum_{j=0}^{N} \hat{f}_j \,\left(\frac{x(t)-\hat{A}_0}{\hat{A}_1}\right)^j  \\ &+   \sum_{j=0}^{N-1} \hat{g}_j \, \left(\frac{x(t)-\hat{A}_0}{\hat{A}_1}\right)^j   x'(t)  \; .
    \label{eq:app1:1ordsys:fg}
\end{aligned}
\end{equation}

A given simulation or experimental measurement of the variables $x(t)$ and $y(t)$ will eventually yield to a set of discrete values \{$x_n$\} and \{$y_n$\}. Suppose that the input-output data has length $L$, so for each value of time $t_n$ with $n\in [0,L-1]$ there is a corresponding value for the dynamical variable $x_n$ and the driven force $y_n$. Notice that in this formulation the time values are not needed to be equally spaced, which presents an advantage compared with the FFT approach given in Ref.~\cite{Gonzalez2023}. Based on these discrete values, Eq.~\ref{eq:app1:1ordsys:fg} can be written in a matrix form as 

\begin{equation}
\begin{bmatrix}
y_0\\y_1\\\vdots \\ y_{L-1}
\end{bmatrix}=\underbrace{\left[\begin{array}{cccccccc}
1 & \left(\frac{x_0-\hat{A}_0}{\hat{A}_1}\right)^1 & \cdots & \left(\frac{x_0-\hat{A}_0}{\hat{A}_1}\right)^N & x'_0 & x'_0 \left(\frac{x_0-\hat{A}_0}{\hat{A}_1}\right)^1 &  \cdots & x'_0 \left(\frac{x_0-\hat{A}_0}{\hat{A}_1}\right)^{N-1}   \\
1 & \left(\frac{x_1-\hat{A}_0}{\hat{A}_1}\right)^1 & \cdots & \left(\frac{x_1-\hat{A}_0}{\hat{A}_1}\right)^N & x'_1 & x'_1 \left(\frac{x_1-\hat{A}_0}{\hat{A}_1}\right)^1 &  \cdots & x'_1 \left(\frac{x_1-\hat{A}_0}{\hat{A}_1}\right)^{N-1}   \\
\vdots & \vdots & \vdots & \vdots & \vdots & \vdots &  \vdots & \vdots \\
1 & \left(\frac{x_{L-1}-\hat{A}_0}{\hat{A}_1}\right)^1 & \cdots & \left(\frac{x_{L-1}-\hat{A}_0}{\hat{A}_1}\right)^N & x'_{L-1} & x'_{L-1} \left(\frac{x_{L-1}-\hat{A}_0}{\hat{A}_1}\right)^1 &  \cdots & x'_{L-1} \left(\frac{x_{L-1}-\hat{A}_0}{\hat{A}_1}\right)^{N-1}   
\end{array}\right]}_{\coloneqq \textbf{A}}
\cdot \begin{bmatrix}
f_0\\f_1\\ \vdots \\ f_N \\g_0\\g_1\\ \vdots \\ g_{N-1}
\end{bmatrix}
\label{eq:app1:matrix}
\end{equation}

By defining the vectors $\vec{y}=[y_0,y_1,\cdots,y_{L-1}]$ and the least-squares regression matrix in the right side as $\textbf{A}$, Eq.~(\ref{eq:app1:matrix}) can be rewritten as

\begin{equation}
    \vec{y}=\textbf{A} \cdot \overrightarrow{\smash{f\!g}\vphantom{f}} \, ,
    \label{eq:formalism:inverse}
\end{equation}

which can be solved for parameters $f_{\!j}$ and $g_{\!j}$ by using the Moore-Penrose inverse\cite{moore1920,penrose1955}

\begin{equation}
 \overrightarrow{\smash{f\!g}\vphantom{f}} = (\textbf{A}^{T}\cdot \textbf{A})^{-1} \cdot \textbf{A}^{T} \cdot \vec{y} \; \; .
\label{eq:formalism:inversefourier}
\end{equation}


\section{SORPS formalism}\label{app:2}

The Sinusoidal Output Response in Power Series (SORPS) formalism is presented in Ref.~\cite{Gonzalez2023}. In this section, we discuss the main results of this theory for completeness. 
In circuit theory and mechanics, it is useful to express a first-order nonlinear system as the following autonomous nonlinear system

\begin{equation}
    y(t)= f(x(t))+g(x(t))x'(t) \, ,
    \label{eq:1ordsys:app}
\end{equation}

where $x(t)$ represents the dynamical variable of the system, and $y(t)$ denotes an external function, referred to as the input and the response of the system, respectively. In circuit theory, the functions $f$ and $g$ are commonly referred to as characteristic curves (CCs) since they define the system. If these curves are known, then the system given by Eq.~(\ref{eq:1ordsys}) is completely defined, and it can be solved using numerical techniques.

Suppose $y(t)$ is piecewise monotone bounded function given by a simulation or experiment where $t$ ranges from 0 to T. We can extend the domain of $y(t)$ to $\mathbb{R}$ by assuming that it repeats periodically with the period $T$. Then, this extension is piecewise monotone bounded periodic function that satisfies Dirichlet conditions\cite{Dirichlet1829} for Fourier series existence, thus having a convergent Fourier series whose value at each point is the arithmetic mean of the left and right limits of the function (for modern expositions of the subject, see Ref.~\cite{Oppenheim1999}). Then, $y(t)$ can be expressed as a Fourier series 

\begin{equation}
    y(t)=a_0+\sum_{k=1}^\infty ( a_k \cos(k\omega t) + b_k \sin(k\omega t) ) \, ,
    \label{eq:fourierv}
\end{equation}

where $\omega=2\pi/T$ represents the fundamental frequency, and the Fourier coefficients are calculated by

\begin{equation}
\begin{aligned}
\begin{cases}
    a_0 &= \frac{\omega}{\pi}\int_{0}^{2\pi/\omega} y(t) dt \\
    a_k &= \frac{\omega}{2\pi}\int_{0}^{2\pi/\omega} y(t) \cos(k\omega t) dt   \, , \,k\geq 1  \\    
    b_k &= \frac{\omega}{2\pi}\int_{0}^{2\pi/\omega} y(t) \sin(k\omega t) dt    \, , \, k\geq 1  \, . 
\end{cases}
\end{aligned}
\end{equation}
       
\begin{equation}        
\begin{aligned}
    y(t)=& \left[ a_0 + \sum_{k=1}^\infty \sum_{\substack{l=0\\ l=l+2 }}^k  \binom{k}{l} \left(1-\sin^2(\omega t)\right)^{l/2}\sin^{k-l}(\omega t) (ab)_{kl} \right]  + \\
     & \left[ \sum_{k=1}^\infty \sum_{\substack{l=1\\ l=l+2 }}^k  \binom{k}{l} \left(1-\sin^2(\omega t)\right)^{(l-1)/2}\sin^{k-l}(\omega t) (ab)_{kl} \right] \cos(\omega t) \; .
\label{eq:outputformalism:sin}
\end{aligned}
\end{equation}

where

\begin{equation}
\begin{aligned}
(ab)_{kl}\coloneqq&  a_k \cos\left(\frac{\pi}{2}(k-l)\right)+b_k \sin\left(\frac{\pi}{2}(k-l)\right) 
=\begin{cases}
(-1)^{(k-l)/2}a_k  \;\;\;\; \text{if } k-l \text{ is even} \\
(-1)^{(k-l-1)/2}b_k  \; \text{if } k-l \text{ is odd} \\
\end{cases}
\label{eq:abkl:def}
\end{aligned}
\end{equation}

The details of this deduction can be found in Ref.~\cite{Gonzalez2023}. 
 Suppose that the dynamical variable $x(t)$ in Eq.~(\ref{eq:1ordsys}) satisfies $x(t)\!=\!A_1\sin(\omega t)+A_0$, where $A_0\in \mathbb{R}$ and $A_1>0$, and $t \in [0,2\pi/\omega]$. Its derivative is given by $x'(t)\!=\!A_1\omega \cos(\omega t)$. 
Subsequently, Eq.~(\ref{eq:outputformalism:sin}) can be expressed as a function of $x(t)$ and its derivative $x'(t)$ as (see Ref.~\cite{Gonzalez2023})

\begin{align}
    y(t)=& \underbrace{\left[ a_0 + \sum_{k=1}^\infty \sum_{\substack{l=0\\ l=l+2 }}^k \sum_{m=0}^{l/2}  \binom{k}{l} \binom{l/2}{m} (-1)^m \left(\frac{x(t)-A_0}{A_1}\right)^{2m+k-l} (ab)_{kl} \right]}_{\coloneqq f\left(\frac{x(t)-A_0}{A_1}\right)}  +\nonumber \\
     &\underbrace{\left[ \sum_{k=1}^\infty \sum_{\substack{l=1\\ l=l+2 }}^k  \sum_{m=0}^{(l-1)/2}  \binom{k}{l} \binom{(l-1)/2}{m} (-1)^m \left(\frac{x(t)-A_0}{A_1}\right)^{2m+k-l} (ab)_{kl} \right] \frac{1}{A_1\omega}}_{\coloneqq g\left(\frac{x(t)-A_0}{A_1}\right)}  \; \; x'(t) \; ,
\label{eq:outputformalism}
\end{align}

By grouping terms with the same power in Eq.~(\ref{eq:outputformalism}), it becomes possible to obtain a Taylor expansion for $f$ and $g$, namely

\begin{align}
    f\!\left( \frac{x(t)-A_0}{A_1}\right)&= \sum_{j=0}^{\infty} f_{\!j} \left( \frac{x(t)-A_0}{A_1}\right)^j  \label{eq:polin:f} \\
    g\!\left( \frac{x(t)-A_0}{A_1}\right) &= \sum_{j=0}^{\infty} g_{\!j} \left( \frac{x(t)-A_0}{A_1}\right)^j
    \label{eq:polin:g}
\end{align}

where 

\begin{align}
    f_{\!j}&=\sum_{\substack{k=j\\ k\leftarrow k+2}}^N [ab]_{kj} \left[ \sum_{\substack{l=k-j\\ l\leftarrow l+2 }}^k \binom{k}{l} \binom{l/2}{(l-k+j)/2} \right] \label{eq:fj_abkj:app} \\
g_{\!j}&= A_1 \omega \sum_{\substack{k=j+1\\ k\leftarrow k+2}}^{N} [ab]_{kj} \left[ \sum_{\substack{l=k-j\\ l\leftarrow l+2 }}^k \binom{k}{l} \binom{(l-1)/2}{(l-k+j)/2} \right] 
\label{eq:gj_abkj:app}
\end{align}

and

\begin{align}
    [ab]_{kj}&\coloneqq \frac{1+(-1)^j}{2}(-1)^{\nicefrac{j}{2}}a_k+\frac{1+(-1)^{j-1}}{2}(-1)^{\nicefrac{(j-1)}{2}}b_k \nonumber\\ &= \begin{cases}
    a_k(-1)^{\nicefrac{j}{2}} & \text{ if } j \text{ is even} \\ 
    b_k(-1)^{\nicefrac{(j-1)}{2}} & \text{ if } j \text{ is odd} \; .
    \end{cases}
    \label{eq:abkj:app}
\end{align}

By the variable change $\hat{k}=k-1$ and $\hat{l}=l-1$, Eq.~\ref{eq:gj_abkj:app} transforms to

\begin{align}
g_{\!j}&=  A_1 \omega \sum_{\substack{\hat{k}=j\\ \hat{k}\leftarrow \hat{k}+2}}^{N-1} [ab]_{(\hat{k}+1)j} \left[ \sum_{\substack{\hat{l}=\hat{k}-j\\ \hat{l}\leftarrow \hat{l}+2 }}^{\hat{k}} \binom{\hat{k}+1}{\hat{l}+1} \binom{\hat{l}/2}{(\hat{l}-\hat{k}+j)/2} \right] \, \; ,
\label{eq:gj_abkj:app:transf}
\end{align}

which is equivalent to the definition of $g_{\!j}$ given in the main text (Eq.~\ref{eq:fgj_abkj}). Besides, Eq.~\ref{eq:fj_abkj:app} corresponds to the definition of $f_{\!j}$ given in the main text.

SORPS method can be expressed as a NARX model (i.e., a particular case of a NARMAX model but without the noise function). This can be visualized by rewriting Eq.~\ref{eq:1ordsys:app} with the finite difference method as

\begin{equation}
x(k)=x(k-1)+\Delta t \frac{y(k-1)-f(x(k-1))}{g(x(k-1))} \; .
\label{eq:narmax:app} 
\end{equation}

The dynamical variable at time $k$ depends on its previous value ($x(k-1)$)  and the previous value of the driven force ($y(k-1)$). A NARX model is defined by\cite{Billings_2013}

\begin{equation}
\begin{aligned}
y(k)= F[&y(k-1),\cdots,y(k-n_y), \\ & u(k-1),\cdots,u(k-n_u)]    
\label{eq:narmax:app:model}
\end{aligned}
\end{equation}
where $n_y$ and $n_u$ are the maximum lag of the output $y$ and the input $u$, respectively. From Eqs.~\ref{eq:narmax:app:model} and \ref{eq:narmax:app}, we identify the dynamical variable as the output and the driven force as the input, and $n_y$=$n_u$=1. Based on this comparison, we propose to call the method as Sinusoidal Output Response in Power Series (SORPS), instead of Sinusoidal Input Response in Power Series (SIRPS) used in Ref.~\cite{Gonzalez2023}.

\section{Matrix formulation}\label{app:3}
In this appendix, we present a matrix formulation of model 2 from Sec.~\ref{sec:formalism:procedure}, which allows us to obtain an alternative matrix formulation of the procedure for system identification. Model 2 can be obtained in a matrix formulation by a procedure with two stages: the first one consist of the transformation of the original power series with hyperparameters $\hat{A}_0$ and $\hat{A}_1$ to a power series with $A_0=0$ and $A_1=1$ (Appendix~\ref{app:3:a}); and the second one consists of the transformation of the power series with $A_0=0$ and $A_1=1$ from the first step to a power series with the desired values of $A_0$ and $A_1$ (Appendix~\ref{app:3:b}). Finally, with these results, model 3 can be obtained in a matrix formulation as shown in Appendix~\ref{app:3:c}. 

\subsection{Power series-based model with hyperparameters $A_0=0$ and $A_1=1$}\label{app:3:a}
In this section, we relate the power series parameters $\hat{f}_j$ and $\hat{g}_j$ from Eq.~\ref{eq:1ordsys:fgtilde} to the polynomial coefficients $F_j$ and $G_j$ defined as follows
\begin{equation}
\begin{aligned}
    f(x(t)) &=\sum_{j=0}^{N} F_j \,\left(x(t)\right)^j  \\
    g(x(t)) &=\sum_{j=0}^{N-1} G_j \, \left(x(t)\right)^j   \; . 
    \label{eq:fgmaclaurin}
\end{aligned}
\end{equation}

By comparing Eqs.~\ref{eq:1ordsys:fgtilde:pow} and \ref{eq:fgmaclaurin}, we obtain
\begin{equation}
\begin{aligned}
    \begin{array}{ll}
   F_j = \displaystyle\sum\limits_{k=j}^N  \binom{k}{j} \frac{\left( -\hat{A}_0\right)^{k-j}}{\hat{A}_1^k} \hat{f}_k &  \text{ for } j \in [0,N]   \\ 
   G_j = \displaystyle\sum\limits_{k=j}^{N-1} \binom{k}{j} \frac{\left( -\hat{A}_0\right)^{k-j}}{\hat{A}_1^k} \hat{g}_k  &  \text{ for } j \in [0,N-1]  \; .
   \end{array}
    \label{eq:fgmaclaurin:a0notnull}
\end{aligned}
\end{equation}
Equation~\ref{eq:fgmaclaurin:a0notnull} can be written in matrix form as

\begin{equation}
\begin{bmatrix}
F_0\\F_1\\\vdots \\ F_N
\end{bmatrix}=\underbrace{\left[\begin{array}{cccccc}
\binom{0}{0}\frac{\left( -\hat{A}_0\right)^{0-0}}{\hat{A}_1^0} & \binom{1}{0}\frac{\left( -\hat{A}_0\right)^{1-0}}{\hat{A}_1^1} & \cdots & \binom{N}{0}\frac{\left( -\hat{A}_0\right)^{N-0}}{\hat{A}_1^N} \\
0 & \binom{1}{1}\frac{\left( -\hat{A}_0\right)^{1-1}}{\hat{A}_1^1} & \cdots & \binom{N}{1}\frac{\left( -\hat{A}_0\right)^{N-1}}{\hat{A}_1^N} \\
\vdots & \vdots & \vdots & \vdots \\
0 & 0 & \cdots & \binom{N}{N}\frac{\left( -\hat{A}_0\right)^{N-N}}{\hat{A}_1^N} 
\end{array}\right]}_{\coloneqq \textbf{M}_{f2F}(\hat{A}_0,\hat{A}_1)}
\cdot \begin{bmatrix}
f_0\\f_1\\ \vdots \\ f_N 
\end{bmatrix}
\label{eq:fgmaclaurin:a0notnull:matf2F}
\end{equation}
\begin{equation}
\begin{bmatrix}
G_0\\G_1\\\vdots \\ G_{N-1}
\end{bmatrix}=\underbrace{\left[\begin{array}{cccccc}
\binom{0}{0}\frac{\left( -\hat{A}_0\right)^{0-0}}{\hat{A}_1^0} & \binom{1}{0}\frac{\left( -\hat{A}_0\right)^{1-0}}{\hat{A}_1^1} & \cdots & \binom{N-1}{0}\frac{\left( -\hat{A}_0\right)^{N-1-0}}{\hat{A}_1^{N-1}} \\
0 & \binom{1}{1}\frac{\left( -\hat{A}_0\right)^{1-1}}{\hat{A}_1^1} & \cdots & \binom{N-1}{1}\frac{\left( -\hat{A}_0\right)^{N-1-1}}{\hat{A}_1^{N-1}} \\
\vdots & \vdots & \vdots & \vdots \\
0 & 0 & \cdots & \binom{N-1}{N-1}\frac{\left( -\hat{A}_0\right)^{N-1-N}}{\hat{A}_1^{N-1}} 
\end{array}\right]}_{\coloneqq \textbf{M}_{g2G}(\hat{A}_0,\hat{A}_1)}
\cdot \begin{bmatrix}
g_0\\g_1\\ \vdots \\ g_{N-1} 
\end{bmatrix} \; \; \; ,
\label{eq:fgmaclaurin:a0notnull:matg2G}
\end{equation}

where the upper triangle matrices are defined as $\textbf{M}_{f2F}(\hat{A}_0,\hat{A}_1)$ and $\textbf{M}_{g2G}(\hat{A}_0,\hat{A}_1)$ for further reference, and they depend on the hyperparameters $\hat{A}_0$ and $\hat{A}_1$. According to linear algebra principles, the inverse of upper triangle matrices exists when all diagonal elements are nonzero- a condition that is always satisfied (note that the case $\hat{A}_1=0$ has been excluded from the beginning of the formalism). It is noteworthy that for the particular case $\hat{A}_0=0$, we obtain a diagonal matrix where the element in the i-th row and i-th column is given by $(\textbf{M})_{ii} =1/\hat{A}_1^i$. 
Equations~\ref{eq:fgmaclaurin:a0notnull:matf2F} and \ref{eq:fgmaclaurin:a0notnull:matg2G} can be compactly written in vector form as

\begin{equation}
    \begin{aligned}
     \vec{F} = \textbf{M}_{f2F}(\hat{A}_0,\hat{A}_1) \cdot \vec{\hat{f}} \\    
      \vec{G} =  \textbf{M}_{g2G}(\hat{A}_0,\hat{A}_1) \cdot \vec{\hat{g}} , \\ 
    \end{aligned}
    \label{eq:fgmaclaurin:fgvecdef}
\end{equation}

where the vectors $\vec{F}$, $\vec{G}$,$\vec{\hat{f}}$ and $\vec{\hat{g}}$ are defined by their corresponding parameters \{$F_j$\}, \{$G_j$\}, \{$\hat{f}_j$\} and \{$\hat{g}_j$\}, ordered as column vectors.

\subsection{Model 2: power series-based model with hyperparameters $A_0$ and $A_1$}\label{app:3:b}
Consider the following power series with values of $A_0$ and $A_1$ that may differ from $\hat{A}_0$ and $\hat{A}_1$, given as
\begin{equation}
\begin{aligned}
    f(x(t)) &=\sum_{j=0}^{N} f_{\!j} \,\left(\frac{x(t)-A_0}{A_1}\right)^j  \\
    g(x(t)) &=\sum_{j=0}^{N-1} g_{\!j} \, \left(\frac{x(t)-A_0}{A_1}\right)^j   \; , 
    \label{eq:1ordsys:fgpol:app}
\end{aligned}
\end{equation}
The results presented in Section~\ref{sec:formalism:resultA} allow us to calculate parameters $f_{\!j}$ and $g_{\!j}$ from $\hat{f}_j$ and $\hat{g}_j$, as explained in the following. Firstly, we can use Eq.~(\ref{eq:fgmaclaurin:fgvecdef}) to compute the $F_j$ and $G_j$ parameters, and secondly, we can utilize its inverse relationship to determine $f_{\!j}$ and $g_{\!j}$. This process can be expressed as 
\begin{equation}
    \begin{aligned}
     \vec{f} &= \textbf{M}_{f2F}^{-1}(A_0,A_1) \cdot \vec{F} \\ 
     \vec{g} &= \textbf{M}_{g2G}^{-1}(A_0,A_1) \cdot \vec{G} \, ,
    \end{aligned}
    \label{eq:fgpol:fgvecdef}
\end{equation}

where $\textbf{M}^{-1}$ symbolizes the inverse of the matrix $\textbf{M}$, and the matrices are evaluated with the hyperparameters $A_0$ and $A_1$ instead of $\hat{A}_0$ and $\hat{A}_1$. Finally, both steps can be combined into a single equation by substituting Eq.~\ref{eq:fgmaclaurin:fgvecdef} into Eq.~\ref{eq:fgpol:fgvecdef},

\begin{equation}
\begin{aligned}
     \vec{f} &= \textbf{M}_{f2F}^{-1}(A_0,A_1) \cdot \textbf{M}_{f2F}(\hat{A}_0,\hat{A}_1) \cdot \vec{\hat{f}} \\ 
     \vec{g} &= \textbf{M}_{g2G}^{-1}(A_0,A_1) \cdot \textbf{M}_{g2G}(\hat{A}_0,\hat{A}_1) \cdot \vec{\hat{g}} \, , 
\end{aligned} 
\label{eq:1ordsys:fghat2fg}
\end{equation}

which is the matrix formulation equivalent to Eq.~\ref{eq:1ordsys:fgexpanded:igualation}

\subsection{Model 3: Fourier analysis-based model matrix formulation}\label{app:3:c}
Eq.~(\ref{eq:fg2abfg}) of the main text can be expressed in a matrix formulation by using Eq.(\ref{eq:1ordsys:fghat2fg}), obtaining
\begin{equation}
    \begin{aligned}
     \overrightarrow{\smash{ab}\vphantom{b}}_{\!f}  &= \textbf{M}_{ab2f}^{-1} \cdot \textbf{M}_{f2F}^{-1}(A_0,A_1) \cdot \textbf{M}_{f2F}(\hat{A}_0,\hat{A}_1) \cdot \vec{\hat{f}} \\    
      \overrightarrow{\smash{ab}\vphantom{b}}_{\!g} &= \frac{1}{A_1 \omega} \textbf{M}_{ab2g}^{-1} \cdot \textbf{M}_{g2G}^{-1}(A_0,A_1) \cdot \textbf{M}_{g2G}(\hat{A}_0,\hat{A}_1) \cdot \vec{\hat{g}} .      
    \end{aligned}    
\end{equation}

\section{Limitations of the SORPS formalism}\label{app:4}

 In this section, we discuss some limitations of the SORPS formalism, starting with the initial motivation from linear first- and second-order systems. 
These considerations not only provide insights into the limitations of the SORPS formalism (especially in its extension to nonlinear higher-order systems), but also shed light on how these limitations could be addressed.

Fourier analysis is a powerful technique extensively applied to linear systems. In the case of a \textbf{l}inear \textbf{t}ime-\textbf{i}nvariant (LTI) system, the response of the system to a single sinusoidal tone is also a sinusoidal tone with the same frequency but eventually differing in amplitude scaling and phase shifting. 
Particularly, for a first-order LTI system, we can obtain the system parameters from the amplitude scaling and phase shifting values or, equivalently, from the Fourier coefficients of the system. 
To be more specific, consider the following first-order LTI system: 

\begin{equation}
 y(t)= a\; x(t) + b\; x'(t) \; ,
 \label{eq:1ordsys:linear}
\end{equation}

where $x(t)$ and $y(t)$ are time-dependent functions, which are commonly called as the dynamical variable and driven force, respectively. Here, $a$ and $b$ are real numbers, and the superscript $'$ denotes the first derivative with respect to time. 
For this LTI system, if we set $x(t)=A\sin(\omega t)$, i.e., a single tone with amplitude $A$ and frequency $\omega$ (with $A$, $\omega>$0), we obtain a single tone with the same frequency for $y(t)$, but eventually with a different amplitude $B$ and phase $\phi$, i.e., $y(t)=B\sin(\omega t+\phi)$. 
Fourier analysis applied to this first-order system allows us to calculate these amplitude scaling $B/A$ and phase shifting $\phi$ based on the Fourier coefficients of $x(t)$ and $y(t)$. 
Specifically, when the dynamical variable is $x(t)=A\sin(\omega t)$, we obtain $y(t)=Aa\sin(\omega t)+A b\omega\cos(\omega t)=B\sin(\omega t+\phi)$, where $B/A=\sqrt{a^2+(b\omega)^2}$ and $\phi=\tan^{-1}(b\omega/a)$.  
From the perspective of the Fourier coefficients, if we consider a time interval with the same value that the period of the single tone (i.e., $T=2\pi/\omega$), we obtain (for $x(t)=A\sin(\omega t)$) the Fourier coefficients $a_0^x=0$, $a_1^x=A$ and $b_1^x=0$, where the superscript $^x$ indicates that are the coefficients for $x(t)$. 
In addition, for $y(t)$, we analogously obtain $a_0^y=0$, $a_1^y=aA$ and $b_1^y=Ab\omega$, which implies that the parameters of the system can be calculated from the Fourier coefficients of $x(t)$ and $y(t)$. Specifically, $a$ is calculated as $a = a_1^y / a_1^x$, and $b$ is determined by $  b = b_1^y /  (a_1^x \omega)$.

This one-to-one correspondence between the Fourier coefficients and the parameters that define the system has motivated the development of the SORPS formalism, which has extended this idea to nonlinear first-order systems\cite{Gonzalez2023}. When applied to nonlinear first-order linear systems, the $a$ and $b$ parameters become functions of the dynamical variable $x(t)$, obtaining 

\begin{equation}
    y(t)= f(x(t))+g(x(t))x'(t) \, ,
    \label{eq:1ordsys:discussion}
\end{equation}

where $f(x)$ and $g(x)$ are the nonlinear characteristic curves of the system. In Ref.~\cite{Gonzalez2023}, a connection was established between the Fourier coefficients of $y(t)$ when  $x(t)=A_0 +A_1 \sin(\omega t)$ and the polynomial fit of $f(x)$ and $g(x)$, where $A_0,A_1\in \mathbf{R}$ and $A_1\neq 0$.  
Consequently, the model of the system, i.e. the parameters that define $f(x)$ and $g(x)$, can be obtained from the Fourier coefficients of $y(t)$ when $x(t)=A_0 +A_1 \sin(\omega t)$, for some hyperparameters $A_0$, $A_1$ and $\omega$.  
A limitation of the methodology employed in Ref. \cite{Gonzalez2023}, is the requirement of a sinusoidal single tone in the dynamical variable $x(t)=A_0 +A_1 \sin(\omega t)$ for system identification purposes.

In this work, we have addressed this limitation by extending the SORPS formalism to obtain system modeling from arbitrary input-output data (i.e., for a given Dataset (DS) composed by the values of $x(t)$ and $y(t)$). 
Thus, parameters of power series-based model that define the systems are obtained through a least-square regression technique, utilizing a power series for $f(x)$ and $g(x)$ with hyperparameters $\hat{A}_0$ and $\hat{A}_1$, as explained in Sec.~\ref{sec:formalism}. Subsequently, these parameters are related, using the SORPS formalism, to the Fourier coefficients of $y(t)$ that would be obtained if a sinusoidal tone were effectively applied to the dynamical variable $x(t)=A_0 +A_1 \sin(\omega t)$, for some specified value of $\omega$.

The first-order linear differential equation of Eq.~\ref{eq:1ordsys:linear}, and its nonlinear version from \ref{eq:1ordsys:discussion} can be applied without further considerations to a wide variety of systems. For example, in electric circuits for the series RL and parallel RC circuits, where in the first case we have $y(t)=v(t)$, $x(t)=i(t)$, $a=R$ and $b=L$, where $R$ and $L$ are the resistance and inductance, $v(t)$ being the total voltage of the circuit, and $i(t)$ being the current. In the second case, we can identify $y(t)=i(t)$, $x(t)=v(t)$, $a=1/R$ and $b=C$, where $C$ is the capacitance, $i(t)$ is the total current of the circuit, and $v(t)$ is the voltage of the resistor or capacitor. 
Another direct application is for the use in viscoelastic materials. 
Based on the Maxwell model, we can identify $y(t)=\epsilon'(t)$, $x(t)=\sigma(t)$, $a=1/\eta$ and $b=1/E$, where $\sigma$ is the stress, $\epsilon$ is the strain, $E$ is the stiffness and $\eta$ is the viscosity of the material. In addition, based on the Kelvin-Voigt model, we can identify $y(t)=\sigma(t)$, $x(t)=\epsilon(t)$, $a=E$ and $b=\eta$ (see Ref.~\cite{Gonzalez2023} and references therein for more details about nonlinear considerations).

Extensions of these equations to higher-order systems can be firstly analyzed with linear systems. 
As known from linear theory the one-to-one correspondence between Fourier coefficients and the parameters defining the system is generally not valid for LTI systems higher than first order. This is mainly due to the underdetermined nature of the system of equations that must be solved (a characteristic that persists even in the linear case). To illustrate this point, consider the following second-order LTI system

\begin{equation}
     y(t)= a\; x(t) + b\; x'(t) + c\; x''(t) \; ,
 \label{eq:2ordsys:linear}
\end{equation}

then, for $x(t)=A\sin(\omega t)$, the resulting expression for $y(t)$ is given by $y(t)=Aa\sin(\omega t)+A b\omega\cos(\omega t)-c \omega^2\cos(\omega t)=A(a-c\omega^2)\sin(\omega t)+Ab\omega\cos(\omega t)$. This leads to the relationships $(a-c\omega^2)=a_1^y/a_1^x$ and $b\omega=b_1^y/a_1^x$. However, this system is underdetermined because $a$ and $c$ can not be uniquely calculated from the Fourier coefficients. The same considerations apply generally to higher-order LTI systems. This suggests that the SORPS formalism is strictly valid only for nonlinear first-order systems, as system equations for higher-order systems are underdetermined.

Despite of this underdetermined nature of the system of equations that must be solved for higher-order nonlinear systems, the SORPS formalism can be potentially applied in a rather indirect way to higher-order systems by rewriting the system as a set of first-order equations. 
We explain this procedure based on a example in the following. Consider a second-order mechanical system based on Eq.~\ref{eq:2ordsys:linear} with a driven force $y(t)$, and involving elastic ($F_k(t)$), viscous ($F_b(t)$) and inertial ($F_A(t)$) forces, we can identify $x(t)$ and $v(t)$ as the position and velocity of the body under study, $c=m$ as the inertial mass, $b$ as the viscosity coefficient, and $a=k$ as the elastic coefficient. Thus, the nonlinear analogous to
Eq.~\ref{eq:2ordsys:linear} applied to this example can be expressed by two different sets of first-order as follows 

\begin{equation}
    \begin{aligned}
    \begin{cases}
        y(t)-F_A(t)&=a(x(t))\; x(t)+b(x(t)) \; x'(t)\\
        F_A(t) &= c(v(t))\; v'(t)  \; ,
    \end{cases}
    \label{eq:2ordsys:linear:2firstord:1}
    \end{aligned}
\end{equation}

and

\begin{equation}
    \begin{aligned}
    \begin{cases}
        y(t)-F_k(t)&=b(v(t))\; v(t)+c(v(t))\; v'(t)\\
        F_k(t)&=a(x(t)) x(t) \; ,
    \end{cases}
    \label{eq:2ordsys:linear:2firstord:2}
    \end{aligned}
\end{equation}

where $v(t)=x'(t)$, and $F_A(t)$ and $F_k(t)$ are time dependent functions that must be known from the simulations or measured in the experiment. 
Thus, SORPS formalism may find applications for systems that can be written as a combination of first-order NDEs as those given by Eqs.~\ref{eq:2ordsys:linear:2firstord:1} and \ref{eq:2ordsys:linear:2firstord:2}.




\end{appendices}


\bibliography{References}

\end{document}